\documentclass[aps,preprint,showpacs,superscriptaddress]{revtex4-2}
\usepackage{amsfonts}  
\usepackage{amssymb}
\usepackage[tbtags]{amsmath}
\usepackage{dsfont}
\usepackage{xcolor}
\usepackage{txfonts}
\usepackage{amssymb}
\usepackage[tbtags]{amsmath}
\usepackage{bm}
\usepackage{xcolor}
\usepackage{graphicx}
\usepackage{ulem}
\begin{document}
\title{Nonhermitian adiabatic perturbation theory  of topological quantization of the average velocity of a magnetic skyrmion
under thermal fluctuations}
\author{Shan-Chang Tang}
\affiliation{Department of Physics  \&  State Key Laboratory of Surface Physics, Fudan University,\\ Shanghai 200433, China}
\author{Yu Shi}
\email{ yushi@fudan.edu.cn}
\affiliation{Department of Physics  \&  State Key Laboratory of Surface Physics,   Fudan University,\\ Shanghai 200433, China}

Phys. Rev. B {\bf 105}, 214415 (2022). 
\begin{abstract}
We study the two-dimensional motion  of a magnetic skyrmion driven by a ratchetlike polarized electric current that is periodic in both space and time. Some general cases are considered, in each of which, in the low temperature and adiabatic limit, regardless of the details of the driving current,  the time and statistical average velocity along any direction is topologically quantized as a Chern number, multiplied by a basic unit.  We make two approaches, one   based on identifying the drift direction, and the other   based on the nonhermitian adiabatic perturbation theory developed for  the Fokker-Planck operator. Both approach applies in the case of periodicity along the direction of the driving current and homogeneity in the transverse direction, for which the analytical result is  confirmed by our numerical simulation on the constituent spins, and a  convenient  experiment is proposed.
\end{abstract}

\maketitle

\section{introduction}

Magnetic skyrmion is a kind of noncollinear spin texture with topological stability,  and  has attracted a lot of interest ever since it had been  theoretically proposed \cite{RosslerBogdanov-2006} and  experimentally  observed  \cite{MuhlbauerBinz-78,YuOnose-161,
YuKanazawa-83,Heinzevon-246,NagaosaTokura-893}. On a large scale, a magnetic skyrmion behaves  as a pointlike object moving on a two-dimensional space. The presence of both driving force and thermal fluctuations motivates us to  consider using magnetic skyrmion to implement the thermal ratchet model, which is an important topic with a wide range of  interests~\cite{reimann}. We proposed to use a magnetic skymion  adiabatically driven by a ratchet-like  spin-polarized electric currents to implement  the adiabatic thermal ratchet\cite{ShiNiu-1596},  in addition to the stochastic force representing thermal fluctuations~\cite{TangShi-2020}, which was the first realization of a thermal ratchet in terms of skyrmions in a uniform temperature. Other realizations of ratchet motions of skyrmions include  unidirectional rotation driven by thermal fluctuations in presence of a temperature gradient \cite{MochizukiYu-44}, Magnus-induced ratchet effects for skyrmions
interacting with asymmetric substrates  \cite{ReichhardtRay-1386,MaReichhardt-2017}, ratchet motion   induced by a biharmonic in-plane magnetic field \cite{ChenLiuJiZheng-2019}.

For the adiabatially driven skyrmion thermal  ratchet \cite{TangShi-2020}, the thermal fluctuations are represented as a stochastic force, and the dynamics is described by  Langevin equation, which was treated by using  Fokker-Planck equation. It was shown that if the driving  electric current is periodic along a specific  direction, which is  different from that of the current itself, and is  determined by a quantity  which is a function of several parameters of the system, in other words,  the periods  along and orthogonal to the direction of the  driving current are  locked in a specific way, then the time and statistical average velocity of the skyrmion is proportional to a closed integral of a curvature of an eigenfunction of an hermitian operator, which is a  similarity transformation of the Fokker-Planck operator. Hence the average velocity is topologically quantized as a Chern number multiplied by a basic unit.  The result implemented a generalization of a one-dimensional adiabatic thermal rathet model \cite{ShiNiu-1596},  and provides a novel method of manipulating magnetic skyrmions. Interesting as it is, this result was  under the special condition concerning the direction of the periodicity, which needs very  careful  arrangement in the experiment.

In this paper, we  make two new approaches and consider several extensions of this problem. First, we make an approach based on identifying the drift direction,  along  which the component of the velocity depends on the driving electric current while independent along the direction  perpendicular to it.  This approach can apply to the case studied in the previous work \cite{TangShi-2020}, which   assumes the periodicity in the drift direction, and is here generalized to a more general form of locking between periods in the direction of the driving current and the orthogonal direction, so that there is  periodicity along the drift direction.

The second approach is based on a perturbation theory for the nonhermitian operator which is a  similarity transformation of the Fokker-Planck operator. This approach applies to the case that the periods along the direction of the driving  electric current and along the orthogonal direction are independent.

These two approaches   both apply to a special case, which is also most  practical,  that  the electric current is periodic along or perpendicular to the direction of the current itself while homogeneous in the direction orthogonal to it. This is because  the homogeneity can be regarded as the periodicity with period $0$. Except for this special case,  the first approach cannot apply to the case that the periods in the longitudinal and transverse directions are independent, as the driving current is then not periodic along the drift direction.

In each of these two approaches, we  find that the time and statistical average velocity is   topologically quantized. For the  special case  in which the driving electric current is periodic along its own direction while homogeneous on the orthogonal direction, we also perform a numerical simulation in terms  of the constituent spins, using the stochastic Landau-Lifschitz-Gilbert equation, confirming the topological quantization. We also propose a convenient  experimental setup for this special case.

On the theoretical aspect, we find a Chern number in a nonhermitian system. From the two previous papers~\cite{ShiNiu-1596,TangShi-2020}, to the present paper, this line of research has been inspired by the analogy   with the adiabatic transport of quantum particle in a periodic potential~\cite{Thouless,NiuThouless-1984}, known as Thouless pump. Indeed, the theoretical framework based on the perturbation theory has been inspired by that for the Thouless pump~\cite{Thouless,NiuThouless-1984}.  But the systems considered in our line of research are classical stochastic systems with thermal fluctuations, rather than quantum fluctuations, as in Thouless pump. It has been noted that  the skyrmions can manifest quantum behavior  at low temperatures~\cite{Ochoa}.  It is interesting, as the future work, to consider the coexistence of quantum and thermal fluctuations and combine elements of Thouless pump and thermal ratchet.

The rest of the paper is organized as the following. In Section \ref{fp}, we introduce the stochastic motion of the  magnetic skyrmion, its description in terms of Langevin and Fokker-Planck equations, and  the similarity transformation of the Fokker-Planck operator.  In Section \ref{app:quantization_diff_vp}, we present  the approach based on identifying the drift direction. In Section \ref{sec:theory},  we develop a nonhermitian perturbation theory based on the eigenfunctions.  Some details are given in Appendices. Especially, we discuss the case  with  independent  periods along and orthogonal to the direction of the  driving current, as well as the special case that the driving electric current is periodic along its own direction while homogeneous on the orthogonal direction, or vice versa. For the  first special case,  we also make the numerical simulation by using the  stochastic Landau-Lifschitz-Gilbert equation, and propose an experiment.  A summary is made in Section \ref{sec:summary}.

\section{Stochastic motion of a magnetic skyrmion \label{fp} }

\subsection{Fokker-Planck Equation and Probability Current  }

Consider a magnetic skyrmion driven by a spin-polarized electric current in a two-dimensional space. At a finite temperature, it is subject to both the driving force and a stochastic force representing the thermal fluctuations.  Its stochastic motion at a finite temperature  can be described in terms of  a Langevin equation with a stochastic term~\cite{TangShi-2020,TroncosoNunez-300}
\begin{equation}
\alpha_d\left[\bm{\dot{q}}-
\frac{\beta}{\alpha}\bm{v}_s\right]
+\alpha_m\bm{\hat{z}}\times[\bm{\dot{q}}
-\bm{v}_s]=\bm{\nu} (t), \label{eqn:skyrmion_langevin}
\end{equation}
where the stochastic variable $\bm{q}=(q_x,q_y)$ represents the position of the skyrmion as a whole, $\alpha_d \equiv \alpha\iint dxdy\left(\frac{\partial\bm{n}}{\partial x}\right)^2$, where $\alpha$ is the Gilbert damping coefficient, $\bm{n}$ represents the direction of each constituent spin,  $\alpha_m \equiv \iint dxdy\bm{n}\cdot\left(\frac{\partial\bm{n}}{\partial x}\times\frac{\partial\bm{n}}{\partial y}\right)$, $\beta$ is the non-adiabatic coefficient, usually  $\beta \ll \alpha$. $\bm{v}_s=-\frac{a^3}{2e}\bm{j}$, where  $\bm{j}$ is the spin-polarized electric current density multiplied by its spin polarization and divided by the magnetic saturation.  $ \bm{\nu} =(\nu_x,\nu_y)$ is the stochastic  force due  to the finite temperature,  satisfying
\begin{equation}
\langle\nu_i(t)\rangle=0, \,
\langle\nu_i(t)\nu_j(t')\rangle
=2\frac{\alpha_dk_BTa^2}{\hbar}
\delta_{ij}\delta(t-t'),
\end{equation}
where $\langle \cdots \rangle $ denotes the statistical ensemble average, $k_B$ is the Boltzmann constant; $\hbar$ is the Planck constant, $T$ is the temperature, $a$ is the lattice constant of the lattice of spins.   For the time being, let us assume that that the electric current is periodic and asymmetric in $x$ and $y$ directions,  with periods $L_x$ and $L_y$ respectively. It is also periodic in time with  period $\mathcal{T}$.

The instantaneous velocity of the skyrmion is $\dot{\mathbf{q}}$. In view of its stochastic nature, we will study its statistical ensemble average $$\langle\dot{\mathbf{q}}\rangle,$$
which is also called particle current~\cite{reimann}.

The statistical nature of the skyrmion  can   be described in terms of the probability density $\rho(\bm{r},t)$, that is,  $\rho(\bm{r},t)dxdy$   is the probability that the skyrmion is located in the region $x\sim x+dx,y\sim y+dy$.
$\rho(\bm{r},t)$ can be obtained as the statistical ensemble average of the constraint that the actual position of the skyrmion $\mathbf{q}(t)$ as a function of $t$,  determined from the Langevin equation,  is  $\mathbf{r}$, that is~\cite{reimann},
 \begin{equation}
\rho(\mathbf{r},t) = \langle \delta(\mathbf{r}-\mathbf{q}(t)\rangle.
\end{equation}
Then from the  Langevin equation, it can be obtained the continuity equation
\begin{equation}
\frac{\partial \rho}{\partial t} + \nabla \cdot {\mathcal J} =0, \label{continuity}
\end{equation}
which is nothing but the continuity equation, with  the probability current density
\begin{equation}
{\mathcal J}(\mathbf{r},t)= \langle  \dot{\mathbf{q}}(t)\delta(\mathbf{r}-\mathbf{q}(t)
\rangle.
\end{equation}
Consequently the total probability current  is
\begin{equation}
  \bm{J}= \iint d^2\bm{r} {\mathcal J} =
\langle \dot{\mathbf{q}} \rangle. \label{jisqdot}
\end{equation}
That is, the probability  current is just the  statistical average of the instantaneous velocity, i.e. the particle current~\cite{reimann}.

From the Langevin equation (\ref{eqn:skyrmion_langevin}), one can derive  the  Fokker-Planck equation~\cite{TangShi-2020,Risken-1585,reimann}
\begin{equation}
-\frac{\partial\rho(\bm{r},t)}{\partial t}={\cal D}\mathcal{O}\rho(\bm{r},t),\label{eqn:skyr_fok_plk}
\end{equation}
where
\begin{equation}
  {\cal D} \equiv \frac{\alpha_d k_B Ta^2}{\hbar(\alpha_m^2+\alpha_d^2)}.\label{eqn:diffusion_constant}
\end{equation}
is the diffusion coefficient,
\begin{equation}
  \mathcal{O}=-\nabla^2+\frac{\partial}{\partial x}(C_1 v_{sx}+C_2 v_{sy}) +\frac{\partial}{\partial y}(-C_2 v_{sx}+C_1 v_{sy})
\end{equation}
is the Fokker-Planck operator, with
\begin{equation}
  C_1\equiv \hbar\frac{\frac{\beta}{\alpha}\alpha_d^2+\alpha_m^2}{\alpha_d k_B Ta^2},\,\,
  C_2 \equiv \hbar\frac{(\frac{\beta}{\alpha}-1)\alpha_m}{k_B Ta^2}.
\end{equation}

For simplicity, we define a 2-vector $\bm{G}(x,y,t)$, whose components are \begin{equation}
\begin{array}{rl}
G_x&\equiv C_1 v_{sx}+C_2 v_{sy},\\
G_y&\equiv -C_2 v_{sx}+C_1 v_{sy}.
\end{array}
\end{equation}
It is clear that $\bm{G}(x,y,t)$ is periodic in time while it is periodic and asymmetric in the  two space dimensions. Then
\begin{equation}
\mathcal{O}=-\nabla^2+\nabla \cdot\mathbf{G}=
-\nabla \cdot (\nabla-\mathbf{G}).
\end{equation}

The Fokker-Planck equation can be rewritten as the  continuity equation (\ref{continuity}), with a different form of the probability current density
where
\begin{equation}
{\mathcal J}= {\cal D}(\bm{G}-\nabla) \rho.
\end{equation}
Therefore the probability current can be obtained  as
\begin{equation}
  \bm{J}=\iint d^2\bm{r} {\mathcal J}  =\iint d^2\bm{r}{\cal D}(\bm{G}-\nabla)\rho,
  \label{eqn:probability_current_original}
\end{equation}
which we will use in the following.

Consider the eigenfunction $\Psi_n$ of $\mathcal{O}$, with eigenvalue $E_n$,
\begin{equation}
 \mathcal{O}\Psi_n=E_n\Psi_n.
\end{equation}
The real part of each $E_n$  is nonnegative, the smallest one being $E_0=0$~\cite{Risken-1585}.  The corresponding ``ground state'' eigenfunction is  $\Psi_0=\rho_0$, which  satisfies  \begin{equation}\mathcal{O}\rho_0=0.\end{equation}
Hence \begin{equation}\nabla \cdot (\nabla-\mathbf{G})\rho_0 =0,\end{equation}  which implies
\begin{equation}
(\nabla-\mathbf{G})\rho_0=\nabla \times {\mathbf A},
\end{equation}
where ${\mathbf A}$ is some function, and can be chosen to be  ${\mathbf A}=A{\mathbf e}_z$, therefore
\begin{equation}
\nabla\rho_0= \bm{G}\rho_0+
\hat{\bm{e}}_x\partial_yA-\hat{\bm{e}}_y\partial_xA. \end{equation}

Note that the instantaneous eigenfunctions themselves are not  solutions to the time-dependent Fokker-Planck equation,  as the Fokker-Planck  operator ${\cal O}$ itself is time-dependent.   So  $\partial_t \Psi_0(t) \neq 0$  though   $E_0(t)=0$.

\subsection{Similarity Transformation }

We make a similarity transformation, under which each eigenfunction $\Psi_n$ is transformed as
\begin{equation}
\Psi_n \rightarrow \psi_n \equiv \rho_0^{-\frac{1}{2}}\Psi_n, \label{tran1}
\end{equation}
with the eigenvalue $E_n$ unchanged, satisfying
\begin{equation}
  \tilde{\mathcal{O}}\psi_n=E_n\psi_n,
\end{equation}
where
\begin{equation}
\tilde{\mathcal{O}}\equiv
  \rho_0^{-1/2}\mathcal{O}\rho_0^{1/2}
\end{equation}
is the transformed operator. For the ``ground state''  $\Psi_0=\rho_0$, the transformed eigenfunction is
\begin{equation}
\psi_0 \equiv \rho_0^{-\frac{1}{2}}\Psi_0= \rho_0^{\frac{1}{2}}.
\end{equation}
Therefore, the similarity transformation can rewritten as
\begin{equation}
  \begin{split}
  \psi_n &\equiv \rho_0^{-\frac{1}{2}}\Psi_n,\\
      \tilde{\mathcal{O}}&\equiv
  \psi_0^{-1}\mathcal{O}\psi_0\\
    &=-\nabla^2-\rho_0^{-1}(\hat{\bm{e}}_x\partial_yA-\hat{\bm{e}}_y\partial_xA)\cdot\nabla+U, \label{eqn:trans_oper}
  \end{split}
\end{equation}
with
\begin{equation}
U\equiv-\Psi_0^{-1}\nabla^2\Psi_0+
\nabla\cdot\bm{G}+\Psi_0^{-1}
\bm{G}\cdot(\nabla\Psi_0).
\end{equation}

Note that when $A$ is independent of $x$ and $y$, $\tilde{\mathcal{O}}$ becomes hermitian, as in our previous work, otherwise,  $\tilde{\mathcal{O}}$    is nonhermitian.

Let us define, in general,
\begin{equation}
\psi \equiv \rho_0^{-\frac{1}{2}}\rho,\label{eqn:psi_definition}
\end{equation}
of which (\ref{tran1}) is the case for eigenfunctions.
The Fokker-Planck equation \eqref{eqn:skyr_fok_plk} can be rewritten in terms of $\Psi$ and $\tilde{\mathcal{O}}$,
\begin{equation}
  -\frac{\partial\psi}{\partial t}=\left({\cal D}\tilde{\mathcal{O}}+\frac{\partial\ln\psi_0}{\partial t}\right)\psi. \label{eqn:trans_fok_plk}
\end{equation}

By substituting \eqref{eqn:psi_definition} into \eqref{eqn:probability_current_original},
the probability current $\bm{J}$ can be obtained as
\begin{equation}
 \bm{J}=-2{\cal D}\iint d^2\bm{r}\rho_0^{\frac{1}{2}}
 \left[\nabla+\frac{1}{2}\rho_0^{-1}
 (\hat{\bm{e}}_x\partial_yA-\hat{\bm{e}}_y
 \partial_xA)\right]\psi,
 \label{eqn:prob_current}
\end{equation}

\section{ Approach based on identifying the drift direction }
\label{app:quantization_diff_vp}

Let's use the orthogonal coordinate system with the direction of the electric current as the $x$ direction, and the direction  orthogonal  to it as the $y$ direction. The initial position of the skyrmion is the origin. From the  Langevin equation \eqref{eqn:skyrmion_langevin}, one can obtain
\begin{align}
\dot{q}_x&=\frac{\frac{\beta}{\alpha}\alpha_d^2+
\alpha_m^2}{\alpha_d^2+\alpha_m^2}v_{sx}(q_x,q_y,t)+
\frac{\alpha_d}{\alpha_d^2+\alpha_m^2}\nu_x+\frac{\alpha_m}{\alpha_d^2+\alpha_m^2}\nu_y,\\
\dot{q}_y&=\frac{(-\frac{\beta}{\alpha}+1)\alpha_d\alpha_m}{\alpha_d^2+\alpha_m^2}v_{sx}(q_x,q_y,t)+\frac{-\alpha_m}{\alpha_d^2+\alpha_m^2}\nu_x+\frac{\alpha_d}{\alpha_d^2+\alpha_m^2}\nu_y.
\end{align}

It is easy to find the direction $v$ in which the velocity component $q_v$ is independent of the driving electric current, and the  orthogonal direction  $u$,
\begin{equation}
  \begin{pmatrix}u\\ v\end{pmatrix}=W\begin{pmatrix}x\\ y\end{pmatrix}.\label{eqn:u-v_coordinate_transformation}
\end{equation}
\begin{equation}
\begin{pmatrix}q_u\\q_v\end{pmatrix}=W\begin{pmatrix}q_x\\q_y\end{pmatrix},\label{eqn:transform}
\end{equation}
where
\begin{equation}
  W=\frac{1}{\sqrt{1+\kappa^2}}\begin{pmatrix}1&-\kappa\\ \kappa&1\end{pmatrix},
\end{equation}
and
\begin{equation}
  \kappa=\frac{(\frac{\beta}{\alpha}-1)\alpha_d\alpha_m}{\frac{\beta}{\alpha}\alpha_d^2+\alpha_m^2}.
\end{equation}

In $uv$ coordinate system, the Langevin equations read
\begin{align}
  \dot{q}_u&={\cal D}\zeta\sqrt{1+\kappa^2}v_{sx}\left(\frac{1}{\sqrt{1+\kappa^2}}q_u+\frac{\kappa}{\sqrt{1+\kappa^2}}q_v,\frac{-\kappa}{\sqrt{1+\kappa^2}}q_u
  +\frac{1}{\sqrt{1+\kappa^2}}q_v,t\right)+\frac{\alpha_d}{\alpha_d^2+\alpha_m^2}\nu_u+\frac{\alpha_m}{\alpha_d^2+\alpha_m^2}\nu_v,\label{eqn:langevin_qu}\\
  \dot{q}_v&=0+\frac{-\alpha_m}{\alpha_d^2+\alpha_m^2}\nu_u+\frac{\alpha_d}{\alpha_d^2+\alpha_m^2}\nu_v,\label{eqn:langevin_qv}
\end{align}
where $\zeta\equiv \hbar\frac{\frac{\beta}{\alpha}\alpha_d^2+\alpha_m^2}{\alpha_d k_B Ta^2}$,
\begin{equation}
  \begin{pmatrix}\nu_u\\ \nu_v\end{pmatrix}=W\begin{pmatrix}\nu_x\\ \nu_y\end{pmatrix}.
\end{equation}
It is clear that
\begin{equation}
  \langle\dot{q}_v\rangle=0, \label{eqn:aver_vel_v}
\end{equation}
as there is no driving term in \eqref{eqn:langevin_qv},
So we can approximately omit the thermal drift in $v$ direction, setting
\begin{equation}
  q_v(t)\approx q_v(0)=0,
\end{equation}
and call $u$ direction as the drift direction.  Then  Eq.~\eqref{eqn:langevin_qu} becomes
\begin{equation}
  \dot{q}_u\approx{\cal D}\zeta\sqrt{1+\kappa^2}v_{sx}\left(\frac{1}{\sqrt{1+\kappa^2}}q_u,\frac{-\kappa}{\sqrt{1+\kappa^2}}q_u,t
  \right)+\frac{\alpha_d}{\alpha_d^2+\alpha_m^2}\nu_u+\frac{\alpha_m}{\alpha_d^2+\alpha_m^2}\nu_v.\label{eqn:langevin_qu_new}
\end{equation}
From Eq.~\eqref{eqn:langevin_qu_new} and Eq.~\eqref{eqn:langevin_qv}, one can obtain the Fokker-Planck equation \cite{Risken-1585}
\begin{equation}
  -\frac{\partial\rho(u,v,t)}{\partial t}={\cal D}\left\{-\nabla^2\rho(u,v,t)
+\zeta\sqrt{1+\kappa^2}\left[\frac{\partial}{\partial u}v_{sx}\left(\frac{1}{\sqrt{1+\kappa^2}}u,\frac{-\kappa}{\sqrt{1+\kappa^2}}u,t
  \right)\rho(u,v,t)\right]\right\}.\label{eqn:skyr_fok_plk_approx}
\end{equation}
Then we separate the variables as
\begin{equation}
  \rho(u,v,t)=\rho_1(u,t)\rho_2(v).
\end{equation}
Consequently, we can obtain two equations for $\rho_1$ and $\rho_2$ respectively
\begin{align}
  &-\frac{\partial\rho_1(u,t)}{\partial t}=-{\cal D}\frac{\partial^2\rho_1(u,t)}{\partial u^2}+{\cal D}\zeta\sqrt{1+\kappa^2}\frac{\partial}{\partial u}\left[v_{sx}\left(\frac{1}{\sqrt{1+\kappa^2}}u,\frac{-\kappa}{\sqrt{1+\kappa^2}}u,t
  \right)\rho_1(u,t)\right]+\lambda\rho_1(u,t),\\
  &\frac{d^2\rho_2(v)}{dv^2}+\frac{\lambda}{{\cal D}}\rho_2(v)=0,
\end{align}
where $\lambda$ is an arbitrary constant.
The second equation is not important since we have already obtained the average velocity along the $v$ direction \eqref{eqn:aver_vel_v}. The first equation can be made simpler by defining $\rho_1'\equiv \rho_1 e^{\lambda t}$,
\begin{equation}
  -\frac{\partial\rho_1'(u,t)}{\partial t}={\cal D}\left[-\frac{\partial^2}{\partial u^2}+\frac{\partial}{\partial u}\zeta\sqrt{1+\kappa^2}v_{sx}\left(\frac{1}{\sqrt{1+\kappa^2}}u,\frac{-\kappa}{\sqrt{1+\kappa^2}}u,t
  \right)\right]\rho_1'(u,t),
\end{equation}
which is the same as the  Fokker-Planck equation for the one-dimensional adiabatic particle transport in a periodic ratchet potential \cite{ShiNiu-1596}.

We now suppose  the space period of $v_{sx}$ is periodic along  $u$ direction, with period  $\mathcal{L}$ and time period $\mathcal{T}$. Then the average velocity of the skyrmion along the $u$ direction is
\begin{equation}
  \overline{\langle\dot{q}_u\rangle}=\mathcal{C}\frac{\mathcal{L}}{\mathcal{T}}.\label{eqn:aver_vel_u}
\end{equation}
where $\mathcal{C}$ is the Chern number. From the average velocity along the $u$ and $v$ direction \eqref{eqn:aver_vel_u} and \eqref{eqn:aver_vel_v}, we can obtain that along the $x$ and $y$ direction
\begin{align}
  &\overline{\langle\dot{q}_x\rangle}=\frac{1}{\sqrt{1+\kappa^2}}\mathcal{C}\frac{\mathcal{L}}{\mathcal{T}},
  \label{x}\\
  &\overline{\langle\dot{q}_y\rangle}=-\frac{\kappa}{\sqrt{1+\kappa^2}}\mathcal{C}\frac{\mathcal{L}}{\mathcal{T}}.
 \label{y}
 \end{align}

This recovers the result in our previous work \cite{TangShi-2020}, where the hermitian condition leads to $v_{sx} = v_{sx}(x-\kappa y,t)$, which means that in $uv$ coordinates,  $v_{sx}$ only depends on $u\equiv (x-\kappa y)/\sqrt{1+\kappa^2}$, while independent of $v\equiv (\kappa x +y)/\sqrt{1+\kappa^2}$, and it was assumed that the period in $u$ is ${\cal L}$.

\eqref{x} and \eqref{y} also apply  to   a generalized case that $v_{sx}=v_{sx}(\kappa_1x+\kappa_2y,t)$ depends on $x$ and $y$ as a function of $\kappa_1x+\kappa_2y$. The hermitian case above is its special case with $\kappa_1=1$ and $\kappa_2=-\kappa$. As a consequence of \eqref{eqn:u-v_coordinate_transformation},
$\kappa_1x+\kappa_2y=\frac{\kappa_1-\kappa_2\kappa}
{\sqrt{1+\kappa^2}}u+\frac{\kappa_2+\kappa_1\kappa}
{\sqrt{1+\kappa^2}}v$.  Hence  the generalized case can be written as $v_{sx}=v_{sx}\left(\frac{\kappa_1-\kappa_2\kappa}
{\sqrt{1+\kappa^2}}u+\frac{\kappa_2+\kappa_1\kappa}
{\sqrt{1+\kappa^2}}v,t\right)$. Now suppose $v_{sx}$ is also periodic in $x$ and $y$, with periods  $L_x$ and $L_y$ respectively.
In order that $v_{sx}=v_{sx}(\kappa_1x+\kappa_2y,t)$ is periodic in $\kappa_1x+\kappa_2y$, it is required that the periods  in  $x$ and $y$ are locked as
\begin{equation}
\kappa_1 L_x=\kappa_2 L_y,  \label{periodicities}
\end{equation}
which is just the period of $v_{sx}$ in $\kappa_1x+\kappa_2y$.
We are now considering the generalized case that
$v_{sx}(\kappa_1x+\kappa_2y,t)$ is periodic in $\kappa_1x+\kappa_2y$ and is approximately independent on $v$.  Remember $v_{sx}(\kappa_1x+\kappa_2y,t)=v_{sx}\left(\frac{\kappa_1-\kappa_2\kappa}
{\sqrt{1+\kappa^2}}u+\frac{\kappa_2+\kappa_1\kappa}
{\sqrt{1+\kappa^2}}v,t\right)$. Hence it is periodic in $\frac{\kappa_1-\kappa_2\kappa}
{\sqrt{1+\kappa^2}}u$ with period $\kappa_1 L_x $. In other words,  it is periodic  in $u$ with period
\begin{equation}
\mathcal{L}=\frac{\kappa_1\sqrt{1+\kappa^2}}
{\kappa_1-\kappa_2 \kappa}L_x.
\end{equation}
Then \eqref{x} and \eqref{y} can be rewritten as
\begin{align}
  &\overline{\langle\dot{q}_x\rangle}=\frac{\kappa_1}{\kappa_1-\kappa_2\kappa}\mathcal{C}\frac{L_x}{\mathcal{T}},\label{eqn:relate_velocity_x}\\
  &\overline{\langle\dot{q}_y\rangle}=-\frac{\kappa_2\kappa}{\kappa_1-\kappa_2\kappa}\mathcal{C}\frac{L_y}{\mathcal{T}}.\label{eqn:relate_velocity_y}
\end{align}

For the case with period-locking but with  \eqref{periodicities} unsatisfied,   $v_{sx}$ is not  periodic in $u$, consequently the present approach does not apply. In general, the present approach applies to all cases in which the driving current is periodic along $u$ direction, including the special case that the driving current is periodic along one of the longitudinal and transverse directions while homogeneous along the other.   This approach does not apply to the case that the periods are independent and both nonzero along these two directions, as the driving current now is not periodic along $u$ direction.

\section{  Approach based on Nonhermitian Adiabatic  Perturbation Theory}

\label{sec:theory}

\subsection{Nonhermitian Adiabatic  Perturbation Theory}

Now we consider another generalization, namely, the case that $\tilde{\mathcal{O}}$ is nonhermitian.

For this purpose, we develop a nonhermitian  adiabatic perturbation theory for Eq.~(\ref{eqn:trans_fok_plk}). First we define the instantaneous eigenfunctions of $\tilde{\mathcal{O}}$. Since $\tilde{\mathcal{O}}$ is not Hermitian, its  eigenfunctions do not necessarily constitute an orthonormal set, that is, $\iint d^2\bm{r} \psi_m^*\psi_n$ is not necessarily equal to  $\delta_{mn}$. Instead, we  define the dual of the original eigenfunctions
\begin{equation}
  \phi_m^*\equiv \sum_l\left(T^{-1}\right)_{ml}
  \psi_l^*, \label{eqn:ortho_partner}
\end{equation}
where $T^{-1}$ is the inverse of $T$, which is defined as $T_{mn} \equiv \langle \psi_m|\psi_n\rangle = \iint d^2\bm{r} \psi_m^*\psi_n$. Clearly,
$$\langle \phi_m|\psi_n\rangle \equiv \iint d^2\bm{r} \phi_m^*\psi_n =\delta_{mn}.$$
It can be easily confirmed that $\phi_m$ is the eigenfunctions of the operator
\begin{equation}
  \tilde{\mathcal{O}}^\dag=-\nabla^2-\nabla\cdot\rho_0^{-1}(\hat{\bm{e}}_x\partial_yA-\hat{\bm{e}}_y\partial_xA)+U,
\end{equation}
with the eigenvalue $E_m^*$ (see Appendix \ref{app:non-Hermi-prop}). Another important relation is
\begin{equation}
\tilde{\mathcal{O}}^\dag\psi_0=0,
\end{equation}
which indicates that $\psi_0$ is also an eigenfunction of $\tilde{\mathcal{O}}^\dag$ with eigenvalue $0$, so we can define   $\phi_0\equiv\psi_0$.

The transformed probability density $\psi$ can then be expanded by the instantaneous eigenfunctions
\begin{equation}
  \psi=\sum_n c_n\psi_n e^{-{\cal D}\int_0^tE_0(t')dt'}.
\end{equation}

Substitute this into the transformed Fokker-Planck equation \eqref{eqn:trans_fok_plk}, calculate the inner products with  $\phi$'s, then we obtain the coefficients through adiabatic perturbation theory. The final result is
\begin{equation}
    \psi=\psi_0+\sum_{n\neq0}\frac{2\langle\phi_n|\dot{\psi}_0\rangle}{{\cal D}(E_0-E_n)}\psi_n.\label{eqn:psi_perturb}
\end{equation}

Now we discuss the adiabatic condition. We consider the case that the potential term dominates the Fokker-Planck operator, that is, the amplitude of $\bm{G}$,
\begin{equation}
  G_0\gg\frac{1}{a},\label{eqn:determinstic_regime}
\end{equation}
$U$ in \eqref{eqn:trans_oper} can also be written as
\begin{equation}
U=\frac{1}{4}(\nabla\ln\rho_0)^2
+\frac{1}{2}(\nabla\cdot\bm{G}).
\end{equation}
Since generically $\rho_0$ is a periodic function, the first term of $U$ possesses a double-well structure. Consequently, the lowest two eigenstates of the system is degenerate approximately with a small eigenvalue difference due to the second term of $U$. Thus the band gap of the system can be estimated to be
\begin{equation}
  \Delta E \sim \frac{G_0}{L},\label{eqn:band_gap_est}
\end{equation}
where $L=\max\{L_x,L_y\}$.
The adiabatic condition is
\begin{equation}
  {\cal T}\gg\frac{1}{{\cal D}\Delta E},\label{eqn:adia_cond}
\end{equation}
where ${\cal T}$ is the time period of the electric current. Hence by substituting \eqref{eqn:band_gap_est} into  \eqref{eqn:adia_cond}, we obtains
\begin{equation}
  {\cal T}\gg\frac{L}{{\cal D}G_0}\label{eqn:adia_cond_final}
\end{equation}

Substituting  \eqref{eqn:psi_perturb} into the probability current \eqref{eqn:prob_current}, we obtain
\begin{equation}
  \bm{J}=-4\sum_{n\neq0}\frac{\langle\phi_0|
  \left[\nabla+\frac{1}{2}
  \rho_0^{-1}(\hat{\bm{e}}_x\partial_yA-
  \hat{\bm{e}}_y\partial_xA)\right]
  \psi_n\rangle\langle\phi_n|
  \dot{\psi}_0\rangle}{E_0-E_n}. \label{originalsummation}
\end{equation}

\subsection{Topologically Quantized Velocity}

$\bm{G}$ is a periodic function, as  a linear combination of the two components of the driving current. For the time being, suppose that periodicities in $x$ and $y$ directions are independent.  So $\psi_n$ and $\phi_n$ must be Bloch waves satisfying
\begin{equation}
  \psi_{n\bm{k}}(\bm{r})=e^{i\bm{k}\cdot\bm{r}}w_{n\bm{k}}(\bm{r}),\phi_{n\bm{k}}(\bm{r})=e^{i\bm{k}\cdot\bm{r}}v_{n\bm{k}}(\bm{r}),
\end{equation}
where $w_{n\bm{k}}(\bm{r})$ and $v_{n\bm{k}}(\bm{r})$ are both periodic functions. The probability current can be regarded as
\begin{equation}
\bm{J}= \bm{J}_{\bm{k}=0},
\end{equation}
where
\begin{equation}
  \bm{J}_{\bm{k}}\equiv -2\sum_{n\neq0}\left(\frac{\langle\phi_{0\bm{k}}|
  \left[\nabla+\frac{1}{2}\rho_0^{-1}
  (\hat{\bm{e}}_x\partial_yA-\hat{\bm{e}}_y
  \partial_xA)\right]|\psi_{n\bm{k}}\rangle\langle\phi_{n\bm{k}}|\dot{\psi}_{0\bm{k}}\rangle}{E_{0\bm{k}}-E_{n\bm{k}}}+\text{c.c.}\right)
  \equiv-2(\bm{J}_{\bm{k}}^h+\bm{J}_{\bm{k}}^{h*}).
\end{equation}
We can rewrite $\bm{J}_{\bm{k}}^h$ as
\begin{equation}
  \bm{J}_{\bm{k}}^h=\frac{i}{2}\langle
  \partial_{\bm{k}}v_{0\bm{k}}|\partial_t w_{0\bm{k}}\rangle-\frac{i}{2}\langle\partial_{\bm{k}}v_{0\bm{k}}|w_{0\bm{k}}\rangle\langle v_{0\bm{k}}|\partial_t w_{0\bm{k}}\rangle,
\end{equation}
the derivation of which is given  in Appendix \ref{app:simp_Jkh}, where it can be seen that  $A$ disappears because it is contained in  the derivative of an operator with respect to $\mathbf{k}$.

If the temperature of the system is very low, the potential term  dominates the transformed Fokker-Planck operator $\tilde{\mathcal{O}}$. As a result, the eigenvalues and the eigenfunctions are insensitive to $\bm{k}$, which means $\psi_{0\bm{k}}\approx \psi_0$ and $\phi_{0\bm{k}}\approx \phi_0$. Thus we obtain  $\psi_{0\bm{k}}\approx \phi_{0\bm{k}}$ and $w_{0\bm{k}} \approx v_{0\bm{k}}$. Consequently  $\bm{J}_{\bm{k}}^h$ can be approximated by
\begin{equation}
  \bm{J}_{\bm{k}}^h\approx \frac{i}{2}\langle\partial_{\bm{k}}w_{0\bm{k}}|\partial_t w_{0\bm{k}}\rangle-\frac{i}{2}\langle\partial_{\bm{k}}w_{0\bm{k}}|w_{0\bm{k}}\rangle\langle w_{0\bm{k}}|\partial_t w_{0\bm{k}}\rangle.
\end{equation}
From this, we can calculate the total probability current
\begin{equation}
  \bm{J}_{\bm{k}}=-i(\langle\partial_{\bm{k}}w_{0\bm{k}}|\partial_t w_{0\bm{k}}\rangle-\langle\partial_t w_{0\bm{k}}|\partial_{\bm{k}} w_{0\bm{k}}\rangle).
\end{equation}

$\bm{J}_{\bm{k}}$ is insensitive to $\bm{k}$, as demonstrated in Appendix \ref{app:insensitive}. It can also be qualitatively understood in the following way. The dependence of $J_{\bm{k}}$ on  $\bm{k}$ mainly originates from the spatial derive in the  Fokker-Planck operator,  which is proportional to temperature, hence is dominated by other terms at low temperatures.

Then the probability current can be written as
\begin{align}
  &J_x(t)\approx\frac{L_x}{2\pi i}\int dk_x(\langle\partial_{k_x}w_{0\bm{k}}|\partial_t w_{0\bm{k}}\rangle-\langle\partial_t w_{0\bm{k}}|\partial_{k_x} w_{0\bm{k}}\rangle),\\
  &J_y(t)\approx\frac{L_y}{2\pi i}\int dk_y(\langle\partial_{k_y}w_{0\bm{k}}|\partial_t w_{0\bm{k}}\rangle-\langle\partial_t w_{0\bm{k}}|\partial_{k_y} w_{0\bm{k}}\rangle).
\end{align}

According to \eqref{jisqdot}, the probability current is just the probabilistic average of the instantaneous velocity of the magnetic skyrmion~\cite{reimann}. Since the driving electric current is periodic in time, the time average of probabilistic average of the velocity of the skyrmion is
\begin{align}
  &\overline{\langle\dot{q}_x\rangle}=\frac{L_x}{\mathcal{T}}\frac{1}{2\pi i}\iint dt dk_x(\langle\partial_{k_x}w_{0\bm{k}}|\partial_t w_{0\bm{k}}\rangle-\langle\partial_t w_{0\bm{k}}|\partial_{k_x} w_{0\bm{k}}\rangle)=\frac{L_x}{\mathcal{T}}\mathcal{C},\label{eqn:velocity_x_final}\\
  &\overline{\langle\dot{q}_y\rangle}=\frac{L_y}{\mathcal{T}}\frac{1}{2\pi i}\iint dt dk_y(\langle\partial_{k_y}w_{0\bm{k}}|\partial_t w_{0\bm{k}}\rangle-\langle\partial_t w_{0\bm{k}}|\partial_{k_y} w_{0\bm{k}}\rangle)=\frac{L_y}{\mathcal{T}}\mathcal{C}'\label{eqn:velocity_y_final}.
\end{align}
where  $\mathcal{C}$ and $\mathcal{C}'$ are Chern numbers. The above expressions clearly demonstrate that the average velocity of a magnetic skyrmion is just a basic unit multiplied by an integer number. This is what we mean by topological quantization. But notice that our  system is a  classical stochastic system.

Notice that he key point is that  $\bm{J}_{\bm{k}}$ is insensitive to $\bm{k}$ at low temperature. It  doesn't really matter whether the velocity is averaged over $k_x$, $k_y$ or the whole Brillouin zone. The result remains unchanged.

If we average the velocity over the whole Brillouin zone, the time and probabilistic average velocity of  the $x$-component velocity  is
\begin{equation}
  \begin{split}
    \overline{\langle\dot{q}_x\rangle}=&\frac{L_y}{2\pi}\int dk_y\frac{L_x}{\mathcal{T}}\frac{1}{2\pi i}\iint dt dk_x(\langle\partial_{k_x}w_{0\bm{k}}|\partial_t w_{0\bm{k}}\rangle-\langle\partial_t w_{0\bm{k}}|\partial_{k_x} w_{0\bm{k}}\rangle)\\
    =&\frac{L_x}{\mathcal{T}}\frac{L_y}{2\pi}\int dk_y\mathcal{C}(k_y).
  \end{split}
\end{equation}

The insensitivity of $\bm{J}_{\bm{k}}$   to $\bm{k}$ implies the insensitivity of $\mathcal{C}(k_y)$ to $k_y$, which is enhanced by the feature that the eigenvalue spectrum  is fully gapped at low temperature and that the Chern number is a topological invariant,  which does not change unless the gap is closed. Thus  $\mathcal{C}(k_y)=\mathcal{C}$ is constant and does not depend on $k_y$. As a result, the average velocity becomes
\begin{equation}
  \begin{split}
    \overline{\langle\dot{q}_x\rangle}=&\frac{L_x}{\mathcal{T}}\left(\frac{L_y}{2\pi}\int dk_y\right)\mathcal{C}\\
    =&\frac{L_x}{\mathcal{T}}\mathcal{C}.
  \end{split}
\end{equation}

One can also start with  \eqref{originalsummation}, with the summation over $n$   replaced as a summation over $n$ and $\bm{k}$, as mentioned by the referee.

Now the probability current can be written as
\begin{equation}
    \bm{J}=-\frac{4}{N_xN_y}\sum_{n\neq0,\bm{k}}\frac{\langle\phi_{0\bm{k}}|\left[\nabla+\frac{1}{2}\rho_0^{-1}(\hat{\bm{e}}_x\partial_yA-\hat{\bm{e}}_y\partial_xA)\right]\psi_{n\bm{k}}\rangle\langle\phi_{n\bm{k}}|\dot{\psi}_{0\bm{k}}\rangle}{E_{0\bm{k}}-E_{n\bm{k}}},
\end{equation}
where $N_x=2\pi/L_x$, $N_y=2\pi/L_y$, $N_x N_y $ is the number of different values of the two-dimensional  discrete crystalline momentum.

Then following the   method similar to above, one can obtain
\begin{equation}
\begin{split}
  \bm{J}=&-\frac{1}{N_xN_y}\sum_{\bm{k}}(-i)(\langle\partial_{\bm{k}}w_{0\bm{k}}|\partial_t w_{0\bm{k}}\rangle-\langle\partial_t w_{0\bm{k}}|\partial_{\bm{k}} w_{0\bm{k}}\rangle)\\
  =&i\frac{L_x}{2\pi}\frac{L_y}{2\pi}\iint dk_x dk_y(\langle\partial_{\bm{k}}w_{0\bm{k}}|\partial_t w_{0\bm{k}}\rangle-\langle\partial_t w_{0\bm{k}}|\partial_{\bm{k}} w_{0\bm{k}}\rangle).
\end{split}
\end{equation}
Consequently, the time-averaged particle current is
\begin{equation}
  \begin{split}
    \overline{ \langle \dot{ \bm{q} } \rangle}=&\frac{1}{\mathcal{T}}\int dt \bm{J}\\
    =&\frac{L_x}{2\pi}\frac{L_y}{2\pi}\frac{1}{\mathcal{T}}\iiint dk_x dk_y dt (\langle\partial_{\bm{k}}w_{0\bm{k}}|\partial_t w_{0\bm{k}}\rangle-\langle\partial_t w_{0\bm{k}}|\partial_{\bm{k}} w_{0\bm{k}}\rangle)
  \end{split}
\end{equation}
Then we again arrive at the conclusion that the time  average of the particle current is topologically quantized.

\subsection{Discussion}

Without loss of generality, suppose that the electric current is along $x$ direction.
We now derive  a constraint on the relation between the two components of the average velocity.
According to the Langevin equation \eqref{eqn:skyrmion_langevin}, we find the following relation
\begin{align*}
  &\overline{\langle\dot{q}_x\rangle}=\frac{\frac{\beta}{\alpha}\alpha_d^2+\alpha_m^2}{\alpha_d^2+\alpha_m^2}\overline{\langle v_{sx}\rangle},\\
  &\overline{\langle\dot{q}_y\rangle}=\frac{\left(-\frac{\beta}{\alpha}+1\right)\alpha_d\alpha_m}{\alpha_d^2+\alpha_m^2}\overline{\langle v_{sx}\rangle}.
\end{align*}

Comparing the above two equation, we conclude that $\overline{\langle\dot{q}_y\rangle}$ is proportional to $\overline{\langle\dot{q}_x\rangle}$, as
\begin{equation}
  \overline{\langle\dot{q}_y\rangle}= -\kappa\overline{\langle\dot{q}_x\rangle},
  \label{eqn:qx_qy_relation}
\end{equation}
where
\begin{equation}
  \kappa \equiv \frac{\left(-\frac{\beta}{\alpha}+1\right)
  \alpha_d\alpha_m}{\frac{\beta}{\alpha}\alpha_d^2
  +\alpha_m^2}. \label{kappa} \end{equation}
This constraint is satisfied by all cases considered in this paper.

In the following, we  consider three subcases. In the first subcase,   the electric current is periodic in $x$ direction while  constant in $y$ direction, which is easy to realize in the experiment, as discussed in Section \ref{sec:experiment}. As a result, the $x$ component of the average velocity  is quantized, as given in   \eqref{eqn:velocity_x_final}, while the argument for the velocity quantization in the preceding section does not apply to  $y$ component. However, it is   obtained from \eqref{eqn:qx_qy_relation} that $\overline{\langle\dot{q}_y\rangle}
  =-\kappa\frac{L_x}{\mathcal{T}}\mathcal{C},$
which is quantized with a more complicated unit. Hence the result for the first case is
\begin{align}
  &\begin{cases}
    &\overline{\langle\dot{q}_x\rangle}=
    \frac{L_x}{\mathcal{T}}\mathcal{C},\\
    &\overline{\langle\dot{q}_y\rangle}=-\kappa \frac{L_x}{\mathcal{T}}\mathcal{C}.
  \end{cases}
  \label{casea}
  \end{align}
This result can also be obtained in the approach based on the drift direction.  Now $v_{sx}$ is independent of $y$, hence  $v_{sx}={\color{red}v_{sx}}\left(\frac{1}{\sqrt{1+\kappa^2}}u,t\right)$. If the period along the $x$ direction is $L_x$, that along the $u$ direction is $\mathcal{L}=L_x\sqrt{1+\kappa^2}$. Substituting  this relation into Eq.~\eqref{x} and Eq.~\eqref{y},   we can reproduce  \eqref{casea}.

In the second case,  the electric current is periodic in the $y$ direction while constant in the $x$ direction.   Consequently the average velocity along the $y$ direction satisfies Eq. \eqref{eqn:velocity_y_final}, while it is the average velocity along the $x$ direction that is obtained from  Eq. \eqref{eqn:qx_qy_relation}, as
$  \overline{\langle\dot{q}_x\rangle} =-\frac{1}{\kappa}\frac{L_y}{\mathcal{T}}\mathcal{C}',
$
which is quantized with a more complicated unit. Hence the result for the second  case is
\begin{align}
  &\begin{cases}
    &\overline{\langle\dot{q}_x\rangle}=-\frac{1}{\kappa} \frac{L_y}{\mathcal{T}}\mathcal{C}',\\
    &\overline{\langle\dot{q}_y\rangle}=\frac{L_y}{\mathcal{T}}\mathcal{C}'.
  \end{cases}
\end{align}
This can also be reproduced  in the approach based on  the drift direction, in a way similar to the first case.

In the  third case, the electric current is periodic in both $x$ and $y$ direction, and the periods are unrelated. This situation 
is difficult to realize in the experiment. Since the relation between the average velocities along the  two directions  satisfy \eqref{eqn:qx_qy_relation}. There are two possibilities,
\begin{align}
  &\begin{cases}
    &\overline{\langle\dot{q}_x\rangle}=\frac{L_x}{\mathcal{T}}\mathcal{C},\\
    &\overline{\langle\dot{q}_y\rangle}=-\kappa \frac{L_x}{\mathcal{T}}\mathcal{C},
  \end{cases}
  \label{casea}
  \end{align}
    or
\begin{align}
  &\begin{cases}
    &\overline{\langle\dot{q}_x\rangle}=-\frac{1}{\kappa} \frac{L_y}{\mathcal{T}}\mathcal{C}',\\
    &\overline{\langle\dot{q}_y\rangle}=\frac{L_y}{\mathcal{T}}\mathcal{C}'.
  \end{cases}
\end{align}
They cannot  be reproduced in the the approach based on the drift direction.

What those Chern numbers are exactly, and which of the two possibilites actually appears in the third case, are determined by the driving electric current.

\subsection{Numerical Simulation}\label{sec:simulation}

In order to confirm our theoretical result, we perform a numerical simulation of the stochastic Landau-Lifschitz-Gilbert equation~\cite{IwasakiMochizuki-87,KongZang-104,LinBatista-1534,MochizukiYu-44,TroncosoNunez-300,TroncosoNunez-1405,Garcia-1597,TangShi-2020}
\begin{equation}
\frac{\partial\bm{n}}{\partial t}+(\bm{v}_s\cdot\nabla)\bm{n}=-\frac{1}{\hbar}\bm{n}\times(\bm{H}_{eff}+\bm{R})+\alpha\bm{n}\times\frac{\partial\bm{n}}{\partial t}+\beta\bm{n}\times(\bm{v}_s\cdot\nabla)\bm{n},\label{eqn:stochastic_LLG}
\end{equation}
which describes the dynamics of the constituent spins of the magnetic skyrmion. $\bm{H}_{eff}\equiv -\frac{\partial\bm{H}_S}{\partial\bm{n}}$ is the effective magnetic field, where the skyrmion Hamiltonian is
\begin{equation}
\bm{H}_S=-J \sum_{\langle ij\rangle}\bm{n}_i\cdot\bm{n}_j-
D\sum_{\langle ij\rangle}\bm{\hat{e}}_{ij}
\cdot{\bm{n}_i\times\bm{n}_j}-
\bm{B}\cdot\sum_i\bm{n}_i-K\sum_in_{iz}^2.
\end{equation}
In this equation, $J$ is the exchange interaction constant, $D$ is the Dzyaloshinskii-Moriya interaction constant \cite{Dzyaloshinsky-1598,Moriya-1599}, $\bm{B}$ is the magnetic field, $K$ is the anisotropic constant, $\bm{R}$ is the random magnetic field, which characterizes the effect of the finite temperature $T$, with
$\langle R_i(\bm{r},t)\rangle=0$, $\langle R_i(\bm{r},t)R_j(\bm{r}',t')\rangle=2\alpha\hbar k_BTa^2\delta_{ij}\delta(\bm{r}-\bm{r}')
\delta(t-t'),$
where $i,j=x,y,z$.

The simulation is performed on a $100\times100$ lattice, which means $L_x=L_y=100$. The Gilbert damping constant is $\alpha=0.1$. The non-adiabatic spin transfer torque constant is $\beta=0$. The Dzyaloshinski-Moriya interaction constant is $D=0.12J$. The magnetic field is $B=0.015J$. The anisotropic energy constant is $K=0.01J$. The electric current density is assumed  to be \cite{Bartussek-1595,TangShi-2020}
\begin{equation}
  \bm{j}=\frac{2e}{a^2\tau}\left[-j_c\left(\cos\frac{2\pi}{L_x}x+\frac{1}{2}\cos\frac{4\pi}{L_x}x\right)-A\cos\frac{2\pi}{\mathcal{T}}t\right]\hat{\bm{e}}_x,\label{eqn:electric_current}
\end{equation}
where $A=0.2$ and $j_c=0.08\sim0.2$,  $\tau\equiv\frac{\hbar}{J}$ is the time unit.  $\bm{j}$ is periodic in $x$ direction while homogeneous in  $y$ direction.  We use the Runge-Kutta method of fourth order, while the time step is chosen to be $0.1\tau$. The choice of the time period $\mathcal{T}$ must satisfy the adiabatic condition \eqref{eqn:adia_cond_final}, under which our adiabatic perturbation theory applies. We have done the simulation for several values of temperature relative to $J/k_B$, given as  $k_BT/J=0.001, 0.01, 0.1$.

According to the definition of $\bm{G}$, the amplitude  $G_0$ of $\bm{G}$,can be approximated as
\begin{equation}
  G_0\sim\frac{\frac{\beta}{\alpha}\alpha_d^2+\alpha_m^2}{\alpha_dk_BTa^2}\hbar\frac{a}{\tau}(j_c+A).
\end{equation}
In our simulation, the corresponding parameters are $\alpha_m=-12.2296\sim10,\alpha_d=1.41767\sim1,j_c+A\sim0.1$, so $G_0$ is approximated by
\begin{equation}
  G_0\sim\frac{1}{a}\left(\frac{k_BT}{J}\right)^{-1}\times10.\label{eqn:G_0_estimate}
\end{equation}

In the deterministic limit, \eqref{eqn:determinstic_regime} must be satisfied, which means $\frac{k_BT}{J}\ll 10$, namely in the low temperature regime. This is actually the case discussed above in the theoretical sections. Then by substituting the expressions in \eqref{eqn:diffusion_constant} and \eqref{eqn:G_0_estimate} for the certain terms in \eqref{eqn:adia_cond_final} and making some approximations, we can get the explicit adiabatic condition
\begin{equation}
  \mathcal{T}\gg10^3\tau.
\end{equation}
Therefore, $\mathcal{T}=5000\tau$ is chosen for the simulation.

From the above parameters, the theoretical values of the two components of the average velocity   can be obtained from (\ref{casea})  as
\begin{align}
  &\overline{\langle\dot{q}_x\rangle}=\mathcal{C}\frac{L_x}{\mathcal{T}}=\mathcal{C}\times0.02\frac{a}{\tau},\label{eqn:x_velocity_explicit}\\
  &\overline{\langle\dot{q}_y\rangle}=\frac{1.41767}{-12.2296}\mathcal{C}\times0.02\frac{a}{\tau}=\mathcal{C}\times\left(-0.00231759\frac{a}{\tau}\right).\label{eqn:y_velocity_explicit}
\end{align}

\begin{figure}[h!]
  \centering
  \includegraphics[scale=0.8]{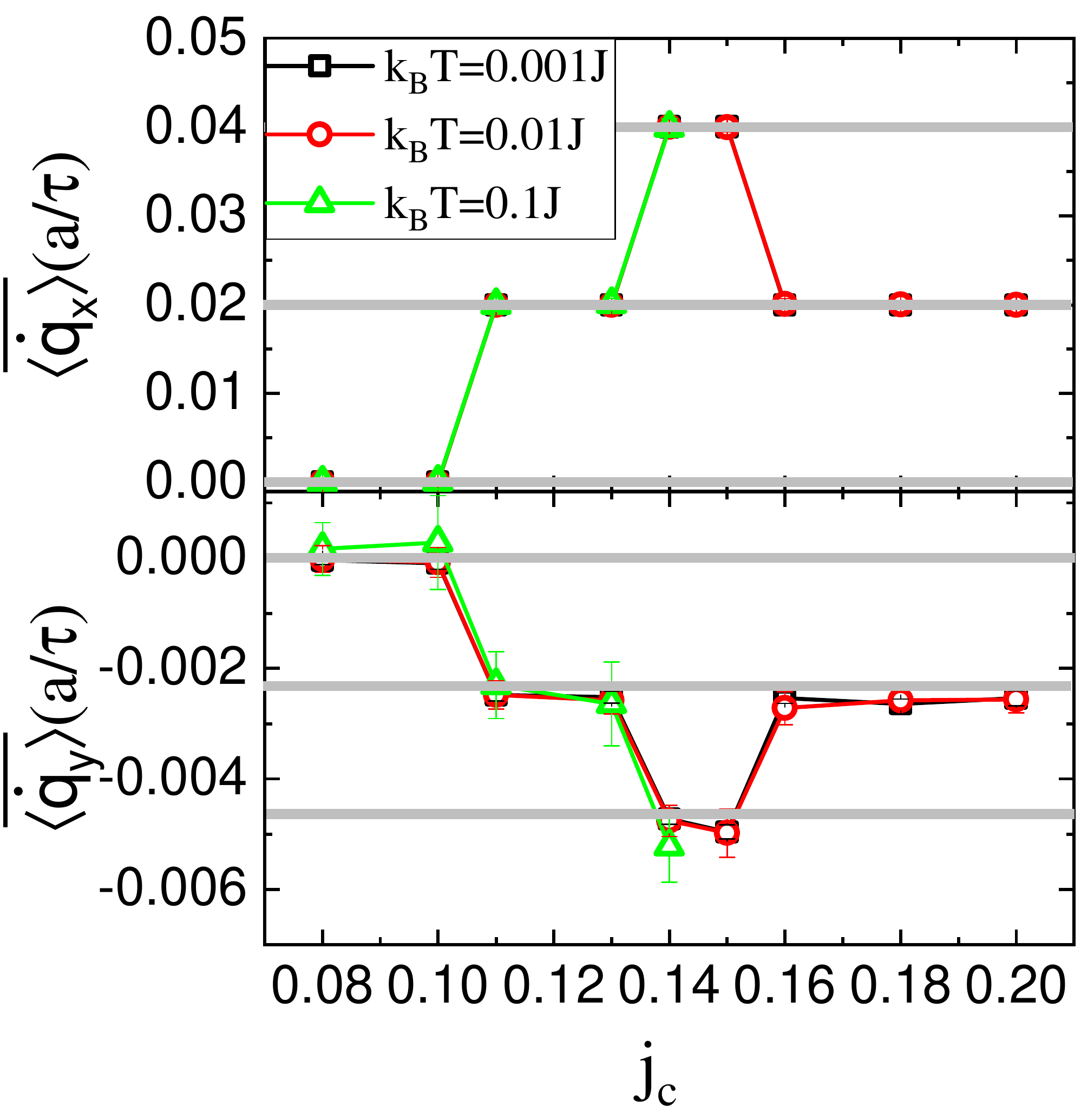}\\
  \caption{The skyrmion's average  velocities  along $x$ and $y$ directions as  functions of the  amplitude $j_c$ of the polarized electric current. The unit of the velocity is  $\frac{a}{\tau}$. Different symbols and colours represent different values of $k_BT$   in unit of $J$. Black curves and squares represent results for $k_BT=0.001J$; red curves and circles represent results for $k_BT=0.01J$; green curves and triangles represent results for  $k_BT=0.1J$. The grey line represents the analytically predicted value of the velocity.}\label{fig:velocity_j_c}
\end{figure}

In the simulation, we obtain  the average velocity, which is averaged over ten periods, versus the parameter $j_c$ for different temperatures, represented as multiplies of exchange interaction constant $J$. The results are  shown in FIG. \ref{fig:velocity_j_c}.

It is clear that the average velocity of the skyrmion  at a low temperature is indeed quantized as given theoretically in \eqref{eqn:x_velocity_explicit} and \eqref{eqn:y_velocity_explicit}.

\subsection{Experimental Proposal}\label{sec:experiment}

\begin{figure}[h!]
  \centering
  \includegraphics[scale=0.3]{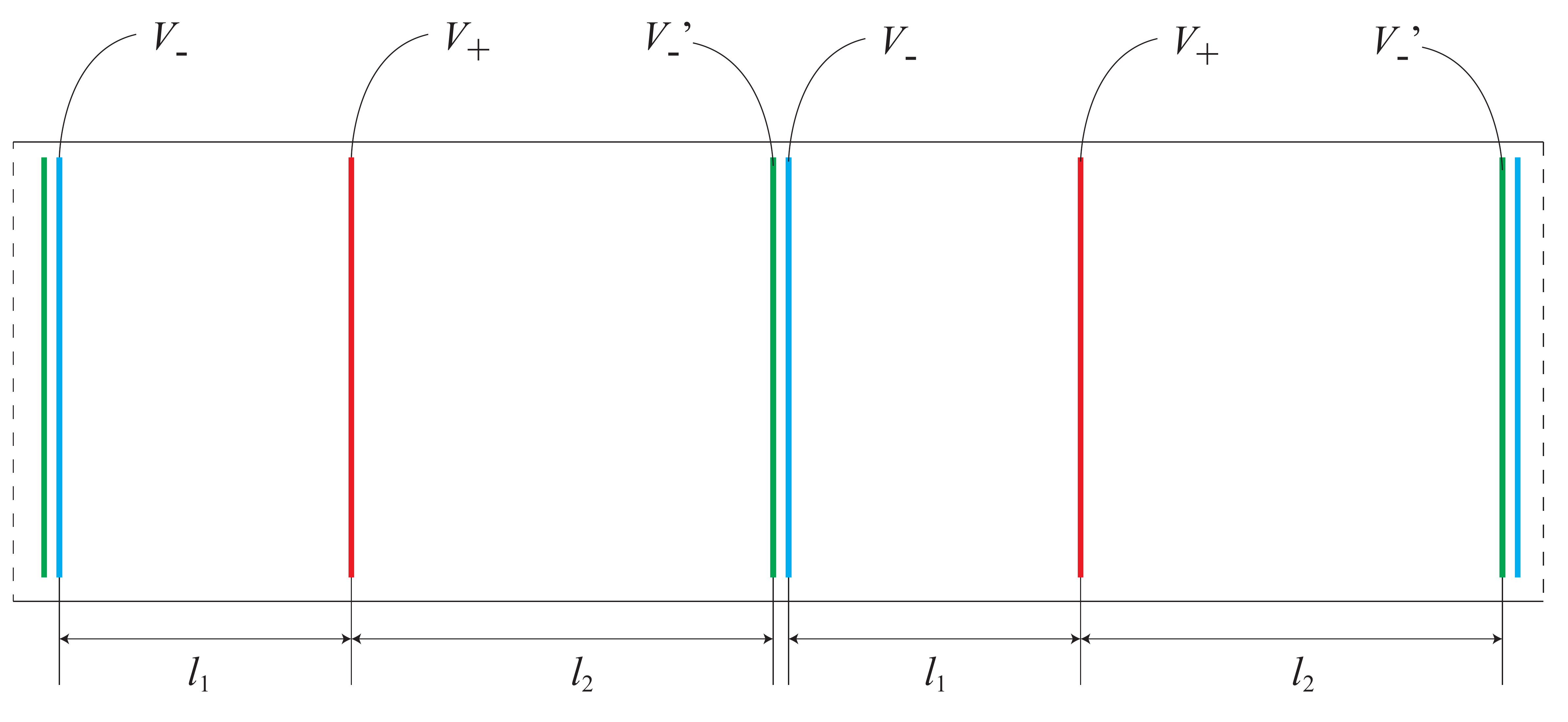}\\
  \caption{The experimental realization of the ratchetlike electric current. The thick lines with different colors represent the electrodes with different electric voltages.}\label{fig:experiment}
\end{figure}

In the above simulation, the electric current density possesses the form \eqref{eqn:electric_current}, which is not easy to realize in the experiment since it is difficult to make the electric current vary with position  as trigonometric functions. However, by using the method we have used in our previous work \cite{TangShi-2020}, we can replace the trigonometric function with the following function
\begin{eqnarray}
  &f(x)=\begin{cases}
    1.5, & 0\leqslant x<20a,\\
    -1  & 20a\leqslant x < 80a,\\
    1.5, & 80a\leqslant x <100a.
  \end{cases}
\end{eqnarray}
Furthermore, $f(x+L_x)=f(x)$. As a result, the electric current density can be written as
\begin{equation}
  \bm{j}=\frac{2e}{a^2\tau}\left[-j_cf(x)-A\cos\frac{2\pi}{\mathcal{T}}t\right]\hat{\bm{e}}_x.\label{eqn:electric_current_exp}
\end{equation}
In order to realize the above ratchetlike electric current, we devise the experiment as shown in Fig. \ref{fig:experiment}. The thick lines are all the electrodes with different electric voltages. The distance between the blue and the electrodes   is $l_1$ while it between the red one and the green one is $l_2$. On the other hand, the distance between the neighboring green and blue electrodes must be as small as possible so that the electric current between them does not affect the motion of the magnetic skyrmion much. In our simulation, $l_1=40a$ and $l_2=60a$. The actual values are not essential.

The red electrodes are all grounded, which means
\begin{equation}
  V_+=0.
\end{equation}
The voltage of each  blue electrode is
\begin{equation}
V_-=-\left(1.5j_c+A\cos\frac{2\pi t}{\cal T}\right)\frac{2e}{a^2\tau} \frac{l_1}{\sigma},
\end{equation}
where  $\sigma$ is the electrical conductivity of the material.   The voltage of each  green electrode is
\begin{equation}
V_-'=-\left(1.0 j_c-A\cos\frac{2\pi t}{\cal T}\right)\frac{2e}{a^2\tau} \frac{l_2}{\sigma}.
\end{equation}
Then the electric current density in different region of the sample is as described by Eq.~\eqref{eqn:electric_current_exp}.

In the actual  experiment, we can first generate a single magnetic skyrmion on the sample where the electrodes are mounted in advance. Then we apply the above electric voltages to the electrodes and the magnetic skyrmion start moving. One measures the change of the position of the skyrmion as a function of time,  from which the instantaneous velocity of the skyrmion can be calculated. Finally, the average velocity of the skyrmion can be obtained by averaging over several periods.

\section{Summary \label{sec:summary} }

We have studied in details the two-dimensional stochastic motion of a magnetic skyrmion driven by a generic  spin-polarized electric current which is periodic 
in time while periodic and asymmetric in the direction of the electric current or in the transverse direction, or in both directions. In any case, the average velocities  along the two directions are shown to be proportional, with the proportional factor given by the drift direction.

We have considered some general cases  significantly beyond the special case considered in our previous work, in which the periods in the  longitudinal and transverse directions  are locked in a special way such that the superposed periodicity is along the drift direction, which is  determined by the parameters of the system.

We have made an approach based on identifying the drift direction, which applies to  a more general case of period-locking, of which the case treated in our previous work is a special one.  If the adiabatic condition is satisfied, the time and probabilistic average of the velocity component along the drift direction is the basic unit, which is the ratio between the space period along this direction and the time period,   multiplied by a Chern number.  The average velocity along the longitudinal and  transverse  directions can be obtained as components. Consequently, the average velocity along any direction, as a projection of that along the drift direction,  is quantized.

We have also made a second approach and  developed  a formalism based on the eigenfunctions of the nonhermitian similarity transformation of the Fokker-Planck operator, and it is assumed that the periods along the longitudinal and transverse  direction  are independent.

In case the driving current is periodic along one of these two direction while homogeneous along the other, the average velocity along this direction is the basic unit  multiplied by a Chern number. Multiplying it by the proportional factor mentioned above gives  the average velocity along the orthogonal  direction.  This result can  be obtained using either of the two approaches.  For the first approach to be applicable, the periods along the longitudinal and transverse directions  should be  in a way that lead to periodicity along the drift direction. This requirement may not be satisfied if the periods along those two directions are independent and both nonzero.

For the  case that the driving current is periodic along its own direction while homogeneous in the transverse direction, we have also performed a numerical simulation which confirms our theoretical prediction, and  have proposed the experimental setup to realize this case,  which is more convenient  than that  in  our previous work  \cite{TangShi-2020}, in which the electric current must be in the form of $f(x-\kappa y)$, where $x$ and $y$ are the spatial coordinates, $\kappa$ is the proportional factor.

The topological quantization provides a  method to robustly manipulate the magnetic skyrmions at a low temperature, which may be useful in memory storage and communication.

\begin{acknowledgments}

We thank Qian Niu for useful discussions. This work was supported by National Science Foundation of China (Grant No. 12075059).
\end{acknowledgments}

\appendix

\section{Eigenfunctions of nonhermitian Operators}\label{app:non-Hermi-prop}

For a nonhermitian operator $O$, define a set of orthogonal basis functions $f_n$, $n=1,2,...$, with
\begin{equation}
\langle f_m|f_n\rangle \equiv\int d\tau f_m^*f_n =\delta_{mn}.
\end{equation}
Then a matrix $O$ can be defined with the matrix elements
\begin{equation}
O_{mn}\equiv \langle f_m|O|f_n\rangle\equiv \int d\tau f_m^* O f_n.
\end{equation}

Suppose $\det{O}\neq 0$, then the matrix can be diagonalized through the similarity transformation
\begin{equation}
P^{-1}OP= diag (E_1,E_2,...) \equiv E,
\end{equation}
where $\{E_n\}$ are eigenvalues.
Thus
\begin{equation}
OP=PE.
\end{equation}
Therefore, the eigenvectors of  $O$ are
\begin{equation}
  a_1=\begin{pmatrix}P_{11}\\P_{21}\\ \vdots\end{pmatrix},a_2=\begin{pmatrix}P_{12}\\P_{22}\\ \vdots\end{pmatrix},\cdots ,a_n=\begin{pmatrix}P_{1n}\\P_{2n}\\ \vdots\end{pmatrix},\cdots,
\end{equation}
with  $E_1$, $E_2$, $\cdots$, $E_n$, $\cdots$. That is,
\begin{equation}
Oa_n=E_na_n.
\end{equation}

On the other hand,
\begin{equation}
P^{-1}O=EP^{-1},
\end{equation}
or
\begin{equation}
O^\dag(P^{-1})^\dag=(P^{-1})^\dag E^\dag,
\end{equation}
which implies that  the eigenvectors of $O^\dag$ are
\begin{equation}
  b_1=\begin{pmatrix}(P^{-1})^\dag_{11}\\(P^{-1})^\dag_{21}\\ \vdots\end{pmatrix},b_2=\begin{pmatrix}(P^{-1})^\dag_{12}\\(P^{-1})^\dag_{22}\\ \vdots\end{pmatrix},\cdots ,b_n=\begin{pmatrix}(P^{-1})^\dag_{1n}\\(P^{-1})^\dag_{2n}\\ \vdots\end{pmatrix},\cdots,
\end{equation}
with eigenvalues $E_1^*$, $E_2^*$, $\cdots$, $E_n^*$, $\cdots$. That is,
\begin{equation}
O^\dag b_n=E_n^\dag b_n.
\end{equation}

It is straightforward to confirm
\begin{equation}
  b_m^\dag a_n=((P ^{-1})^\text{T}_{1m},(P^{-1})^\text{T}_{2m},\cdots)
  \begin{pmatrix}P_{1n}\\P_{2n}\\ \vdots\end{pmatrix}=
  (P^{-1})_{m1}P_{1n}+(P^{-1})_{m2}P_{2n}+\cdots
  =(P^{-1}P)_{mn}=\delta_{mn}.
\end{equation}

The eigenfunctions of the operator $O$ and $O^\dag$ can be obtained as
\begin{align}
  \psi_n=&\sum_i P_{in}f_i,\\
  \phi_n=&\sum_i(P^{-1})^\dag_{in}f_i.
\end{align}
By using Dirac notation, the operator $\hat{A}$ can be written as $$O=\sum_{mn}O_{mn}|f_m\rangle\langle f_n|.$$ Therefore
\begin{align*}
  O|\psi_n\rangle=&\sum_{ml}O_{ml}|f_m\rangle\langle f_l|\sum_i P_{in}|f_i\rangle\\
  =&\sum_{iml}O_{ml}P_{in}\delta_{il}|f_m\rangle=\sum_{ml}O_{ml}P_{ln}|f_m\rangle=\sum_{ml}P_{ml}E_{ln}|f_m\rangle=E_n\sum_mP_{mn}|f_m\rangle=E_n|\psi_n\rangle,\\
  O^\dag|\phi_n\rangle=&\sum_{ml}O^\dag_{ml}|f_m\rangle\langle f_l|\sum_i(P^{-1})^\dag_{in}|f_i\rangle\\
  =&\sum_{iml}O^\dag_{ml}(P^{-1})^\dag_{in}\delta_{il}|f_m\rangle=\sum_{ml}O^\dag_{ml}(P^{-1})^\dag_{ln}|f_m\rangle=\sum_{ml}(P^{-1})^\dag_{ml}E^\dag_{ln}|f_m\rangle=E^*_n\sum_m(P^{-1})_{mn}|f_m\rangle\\
  =&E^*_n|\phi_n\rangle,
\end{align*}
which confirms that $|\psi_n\rangle$ and   $|\phi_n\rangle$ are indeed eigenfunctions of $O$ and $O^\dag$, respectively. Now the inner products of these eigenfunctions can be calculated as
\begin{align}
  \langle\psi_m|\psi_n\rangle=&\sum_{ij}P^*_{im}P_{jn}\langle f_i|f_j\rangle=\sum_{ij}(P^\dag)_{mi}P_{jn}\delta_{ij}=(P^\dag P)_{mn}=T_{mn},\\
  \langle\phi_m|\psi_n\rangle=&\sum_{ij}\left[(P^{-1})^\dag_{im}\right]^*P_{jn}\langle f_i|f_j\rangle=\sum_{ij}(P^{-1})_{mi}P_{jn}\delta_{ij}=(P^{-1}P)_{mn}=\delta_{mn}.
\end{align}
These inner products   help figure out whether the orthogonal partner defined in \eqref{eqn:ortho_partner} is the eigenfunctions of the hermitian conjugate operator obtained here.
\begin{align*}
  \sum_l(T^{-1})_{ml}\psi_l^*=&\sum_{ln}(P^{-1})_{mn}(P^{-1})^\dag_{nl}\sum_iP^*_{il}f_i^*\\
  =&\sum_{iln}(P^{-1})_{mn}(P^{-1})^\dag_{nl}P^\dag_{li}f_i^*=\sum_{in}(P^{-1})_{mn}\delta_{ni}f_i^*=\sum_n(P^{-1})_{mn}f_n^*\\
  =&\left(\sum_n(P^{-1})^\dag_{nm}f_n\right)^*=\phi_m^*.
\end{align*}

\section{Insensitivity of $\bm{J}_{\bm{k}}$ to $\bm{k}$}\label{app:insensitive}

The following Hermitian and antihermitian operators can be obtained from the transformed Fokker-Planck operator \eqref{eqn:trans_oper}
\begin{align}
  L_H=&\frac{\tilde{\mathcal{O}}+\tilde{\mathcal{O}}^\dag}{2}=-\nabla^2+\frac{1}{2}(\nabla\cdot\bm{M})+U,\\
  L_A=&\frac{\tilde{\mathcal{O}}-\tilde{\mathcal{O}}^\dag}{2}=-\bm{M}\cdot\nabla-\frac{1}{2}(\nabla\cdot\bm{M}),
\end{align}
where $\bm{M}\equiv\rho_0^{-1}(\hat{\bm{e}}_x\partial_y A-\hat{\bm{e}}_y\partial_xA)$. Then we can construct an  operator from the above two operators \cite{Risken-1585}
\begin{equation}
  \mathcal{H}=L_H-i\eta L_A=-\nabla^2+i\eta\bm{M}\cdot\nabla+U',
\end{equation}
where $U' \equiv U+\frac{1+i\eta}{2}(\nabla\cdot\bm{M})$.

It can be seen that when $\eta =i$, $\mathcal{H}= \tilde{\mathcal{O}}$. When $\eta$ is real,  $\mathcal{H}$ is  Hermitian.

For the time being, we assume $\eta$ is real.
The eigenfunctions of $\mathcal{H}$ are $\psi'_{n\bm{k}}$, satisfying
\begin{equation*}
  \mathcal{H}(\eta)\psi'_{n\bm{k}}(\eta)=E'_{n\bm{k}}(\eta)\psi'_{n\bm{k}}(\eta).
\end{equation*}
They are of course Bloch wave functions and their periodic parts are $w'_{n\bm{k}}$'s, which satisfy
\begin{equation*}
  \mathcal{H}'(\eta)w'_{n\bm{k}}(\eta)=E'_{n\bm{k}}(\eta)w'_{n\bm{k}}(\eta),
\end{equation*}
where
\begin{equation}
  \mathcal{H}' \equiv -(\nabla+i\bm{k})^2+i\eta\bm{M}\cdot(\nabla+i\bm{k})+U'.
\end{equation}
Then we obtain the probability current
\begin{align}
  \bm{J}'_{\bm{k}}(\eta)=&-i(\langle\partial_{\bm{k}}w'_{0\bm{k}}|\partial_t w'_{0\bm{k}}\rangle-\langle\partial_t w'_{0\bm{k}}|\partial_{\bm{k}} w'_{0\bm{k}}\rangle)=-2(\bm{J}_{\bm{k}}'^h+
  \bm{J}_{\bm{k}}'^{h*}),\end{align}
where
\begin{align}
  \bm{J}_{\bm{k}}'^h(\eta)=&\frac{i}{2}\langle\partial_{\bm{k}}w'_{0\bm{k}}|\partial_t w'_{0\bm{k}}\rangle-\frac{i}{2}\langle\partial_{\bm{k}}w'_{0\bm{k}}|w'_{0\bm{k}}\rangle\langle w'_{0\bm{k}}|\partial_t w'_{0\bm{k}}\rangle. \label{jkh}
\end{align}

Following a method of Niu and Thouless for   the quantized adiabatic charge transport  \cite{NiuThouless-1984}, we first write $\bm{J}'_{\bm{k}}$ in the form of Green functions and then prove its insensitivity to $\bm{k}$ in the following.

From \eqref{jkh},
\begin{equation}
  \begin{split}
    \bm{J}_{\bm{k}}'^h(\eta)=&\frac{i}{2}\langle\partial_{\bm{k}}w'_{0\bm{k}}|\partial_t w'_{0\bm{k}}\rangle-\frac{i}{2}\langle\partial_{\bm{k}}w'_{0\bm{k}}|w'_{0\bm{k}}\rangle\langle w'_{0\bm{k}}|\partial_t w'_{0\bm{k}}\rangle\\
    =&\frac{i}{2}\sum_{n\neq0}\langle\partial_{\bm{k}}w'_{0\bm{k}}|w'_{n\bm{k}}\rangle\langle w'_{n\bm{k}}|\partial_t w'_{0\bm{k}}\rangle\\
    =&-\frac{i}{2}\sum_{n\neq0}\langle w'_{0\bm{k}}|\partial_{\bm{k}}w'_{n\bm{k}}\rangle\langle w'_{n\bm{k}}|\partial_t w'_{0\bm{k}}\rangle.
    \label{eqn:ProbCurrHalf'}
  \end{split}
\end{equation}
Then calculate the derivatives of  both sides of $\mathcal{H}'(\eta)w'_{n\bm{k}}(\eta)=E'_{n\bm{k}}(\eta)w'_{n\bm{k}}(\eta)$ with respect to $\bm{k}$,
\begin{equation}
  \frac{\partial\mathcal{H}'}{\partial\bm{k}}w'_{n\bm{k}} +\mathcal{H}'\frac{\partial w'_{n\bm{k}}}{\partial \bm{k}}
  =\frac{\partial E'_{n\bm{k}}}{\partial\bm{k}}w'_{n\bm{k}} +E'_{n\bm{k}}\frac{\partial w'_{n\bm{k}}}{\partial \bm{k}}.
\end{equation}
The inner product of both sides of the above equation with $w'_{0\bm{k}}$, for  $n\neq 0$, leads to
\begin{equation}
  \begin{split}
    &\langle w'_{0\bm{k}}|\frac{\partial\mathcal{H}'}{\partial\bm{k}}|w'_{n\bm{k}}\rangle +E'_{0\bm{k}}\langle w'_{0\bm{k}}|\frac{\partial w'_{n\bm{k}}}{\partial \bm{k}}\rangle
  =E'_{n\bm{k}}\langle w'_{0\bm{k}}|\frac{\partial w'_{n\bm{k}}}{\partial \bm{k}}\rangle,
  \end{split}
\end{equation}
therefore,
\begin{equation}
  \begin{split} \langle w'_{0\bm{k}}|\frac{\partial w'_{n\bm{k}}}{\partial \bm{k}}\rangle
  =-\frac{1}{E'_{0\bm{k}}-E'_{n\bm{k}}}\langle w'_{0\bm{k}}|\frac{\partial\mathcal{H}'}
  {\partial\bm{k}}|w'_{n\bm{k}}\rangle.
  \end{split}
\end{equation}
which is substituted  into Eq.~\eqref{eqn:ProbCurrHalf'} to obtain
\begin{equation}
\begin{split}
  \bm{J}_{\bm{k}}'^h(\eta)=&\frac{i}{2}\sum_{n\neq0}\frac{\langle w'_{0\bm{k}}|\frac{\partial\mathcal{H}'}{\partial\bm{k}}|w'_{n\bm{k}}\rangle\langle w'_{n\bm{k}}|\partial_t w'_{0\bm{k}}\rangle}{E'_{0\bm{k}}-E'_{n\bm{k}}}\\
  =&\frac{i}{2}\sum_{n\neq0}\frac{\langle w'_{0\bm{k}}|(-2i)(\nabla+i\bm{k}-\frac{i\eta}{2}\bm{M}) |w'_{n\bm{k}}\rangle\langle w'_{n\bm{k}}|\partial_t w'_{0\bm{k}}\rangle}{E'_{0\bm{k}}-E'_{n\bm{k}}}\\
  =&\sum_{n\neq0}\frac{\langle \psi'_{0\bm{k}}|(\nabla-\frac{i\eta}{2}\bm{M}) |\psi'_{n\bm{k}}\rangle\langle \psi'_{n\bm{k}}|\partial_t \psi'_{0\bm{k}}\rangle}{E'_{0\bm{k}}-E'_{n\bm{k}}},
  \label{eqn:ProbCurrHalf'_med}
\end{split}
\end{equation}
where we have used  $$\frac{\partial\mathcal{H}'}{\partial\bm{k}}
=-2i(\nabla+i\bm{k}-\frac{i\eta}{2}\bm{M}),$$ and $$\psi_{n\bm{k}}=e^{i\bm{k}\bm{x}}w_{n\bm{k}}.$$
Since $\mathcal{H}\psi'_{n\bm{k}}=E'_{n\bm{k}}\psi'_{n\bm{k}}$, one obtains
\begin{equation}
  \begin{split}
    &\left(\nabla-\frac{i\eta}{2}\bm{M}\right)(\mathcal{H}\psi'_{n\bm{k}})
    =\left(\nabla-\frac{i\eta}{2}\bm{M}\right)
    (E'_{n\bm{k}}\psi'_{n\bm{k}}),
\end{split}
\end{equation}
Thus
 \begin{equation}
  \begin{split}
    &\left[\nabla-\frac{i\eta}{2}\bm{M},\mathcal{H}\right]\psi'_{n\bm{k}}+\mathcal{H}\left(\nabla-\frac{i\eta}{2}\bm{M}\right)\psi'_{n\bm{k}}
    =E'_{n\bm{k}}\left(\nabla-\frac{i\eta}{2}
    \bm{M}\right)\psi'_{n\bm{k}},
\end{split}
\end{equation}
Therefore
 \begin{equation}
  \begin{split}
&\langle\psi'_{0\bm{k}}|\left[\nabla-\frac{i\eta}{2}\bm{M},\mathcal{H}\right]|\psi'_{n\bm{k}}\rangle+\langle\psi'_{0\bm{k}}|\mathcal{H}\left(\nabla-\frac{i\eta}{2}\bm{M}\right)|\psi'_{n\bm{k}}\rangle
    =E'_{n\bm{k}}\langle\psi'_{0\bm{k}}|\left(\nabla-\frac{i\eta}{2}\bm{M}\right)|\psi'_{n\bm{k}}\rangle,
\end{split}
\end{equation}
Hence
 \begin{equation}
  \begin{split}       &\langle\psi'_{0\bm{k}}|\left(\nabla-\frac{i\eta}{2}\bm{M}\right)|\psi'_{n\bm{k}}\rangle
    =\frac{\langle\psi'_{0\bm{k}}|\left[\mathcal{H},\nabla-\frac{i\eta}{2}\bm{M}\right]|\psi'_{n\bm{k}}\rangle}{E'_{0\bm{k}}-E'_{n\bm{k}}}. \label{eq1}
  \end{split}
\end{equation}

On the other hand,
\begin{equation}
  \begin{split}
    &\dot{\mathcal{H}}
    \psi'_{0\bm{k}}+\mathcal{H}\partial_t
    \psi'_{0\bm{k}}=\dot{E}'_{0\bm{k}}
    \psi'_{0\bm{k}}+E'_{0\bm{k}}\partial_t
    \psi'_{0\bm{k}}.
\end{split}
\end{equation}
thus
 \begin{equation}
  \begin{split}
&\langle\psi'_{n\bm{k}}|\dot{\mathcal{H}}|\psi'_{0\bm{k}}\rangle+\langle\psi'_{n\bm{k}}|\mathcal{H}|\partial_t\psi'_{0\bm{k}}\rangle =\dot{E}'_{0\bm{k}}\langle\psi'_{n\bm{k}}|\psi'_{0\bm{k}}\rangle+E'_{0\bm{k}}\langle\psi'_{n\bm{k}}|\partial_t\psi'_{0\bm{k}}\rangle,
\end{split}
\end{equation}
therefore
 \begin{equation}
  \begin{split}
 &\langle\psi'_{n\bm{k}}|\dot{\mathcal{H}}|\psi'_{0\bm{k}}\rangle+E'_{n\bm{k}}\langle\psi'_{n\bm{k}}|\partial_t\psi'_{0\bm{k}}\rangle =E'_{0\bm{k}}\langle\psi'_{n\bm{k}}|\partial_t\psi'_{0\bm{k}}\rangle,
\end{split}
\end{equation}
hence
 \begin{equation}
  \begin{split}   &\langle\psi'_{n\bm{k}}|\partial_t\psi'_{0\bm{k}}\rangle
    =\frac{\langle\psi'_{n\bm{k}}|\dot{\mathcal{H}}|\psi'_{0\bm{k}}\rangle}{E'_{0\bm{k}}-E'_{n\bm{k}}}.
\label{eq2}
  \end{split}
\end{equation}
Substituting \eqref{eq1} and \eqref{eq2}  into Eq.~\eqref{eqn:ProbCurrHalf'_med}, one obtains
\begin{equation}
    \bm{J}_{\bm{k}}'^h(\eta)=\sum_{n\neq0}\frac{\langle\psi'_{0\bm{k}}|[\mathcal{H},\nabla-\frac{i\eta}{2}\bm{M}]|\psi'_{n\bm{k}}\rangle \langle\psi'_{n\bm{k}}|\dot{\mathcal{H}}|\psi'_{0\bm{k}}\rangle}{(E'_{0\bm{k}}-E'_{n\bm{k}})^3}.
\end{equation}

According to the residue theorem, $$\frac{1}{(E'_{0k}-E'_{nk})^3}=-\frac{1}{2}\ointctrclockwise_\mathcal{C}\frac{dz}{2\pi i}\frac{1}{(z-E'_{0q})^2}\frac{1}{(z-E'_{nq})^2}, $$ where the path $\mathcal{C}$ encircles $E'_{0q}$. Inserting this into the above equation, we obtain
\begin{equation}
  \bm{J}_{\bm{k}}'^h(\eta)=-\frac{1}{2}\ointctrclockwise_\mathcal{C}\frac{dz}{2\pi i}Tr\left[P_0g\left[\mathcal{H},\nabla-\frac{i\eta}{2}\bm{M}\right]gP_Eg\dot{\mathcal{H}}g\right],
\end{equation}
where $g\equiv\frac{1}{z-\mathcal{H}}$ and $P_0\equiv|\psi'_{0\bm{k}}\rangle\langle\psi'_{0\bm{k}}|,P_E\equiv\sum_{n\neq 0}|\psi'_{n\bm{k}}\rangle\langle\psi'_{n\bm{k}}|$. In the same way, we obtain
\begin{equation}
   \bm{J}_{\bm{k}}'^{h*}(\eta)=-\frac{1}{2}\ointctrclockwise_\mathcal{C}\frac{dz}{2\pi i}Tr\left[P_Eg\left[\mathcal{H},\nabla-\frac{i\eta}{2}\bm{M}\right]gP_0g\dot{\mathcal{H}}g\right].
\end{equation}
It is also straightforward to prove that $$\ointctrclockwise_\mathcal{C}\frac{dz}{2\pi i}
\left(Tr\left[P_0g\left[\mathcal{H},\nabla-\frac{i\eta}{2}\bm{M}\right]gP_0g\dot{\mathcal{H}}g\right]+Tr\left[P_Eg\left[\mathcal{H},\nabla-\frac{i\eta}{2}\bm{M}\right]gP_Eg\dot{\mathcal{H}}g\right]\right)=0,$$
by using the residue theorem. Therefore, we can obtain the probability current for $\mathcal{H}$
\begin{equation}
  \begin{split}
    \bm{J}'_{\bm{k}}(\eta)=
    &\ointctrclockwise_\mathcal{C}\frac{dz}{2\pi i}Tr\left[g\left[\mathcal{H},
    \nabla-\frac{i\eta}{2}\bm{M}\right]gg
    \dot{\mathcal{H}}g\right]\\
    =&\ointctrclockwise_\mathcal{C}\frac{dz}{2\pi i}Tr\left[\left[g,\nabla-\frac{i\eta}{2}\bm{M}\right]g\dot{\mathcal{H}}g\right].\label{eqn:app_prob_curr}
  \end{split}
\end{equation}
Because $\nabla-\frac{i\eta}{2}\bm{M}=-\frac{1}{2}[\mathcal{H},\bm{x}]$, we can further simplify $\bm{J}'_{\bm{k}}(\eta)$. However, the inclusion of the $\bm{x}$ operator in the integrand makes the integration not well defined, since it   diverges if the size of the system is infinite. Consequently, we modify $\bm{x}$ so that it is periodic. Suppose the periods along the $x$ and $y$ direction are both $L$. We now define the operator $\bm{\xi}$, with its components satisfying
\begin{equation}
  \xi_i=x_i-L\theta(x_i)+\frac{L}{2},
\end{equation}
where $x_1=x$, $x_2=y$, $\theta(x)$ is the standard Heaviside function. Thus  in the region $-\frac{L}{2}<\xi_i\leqslant\frac{L}{2}$.   Then we can obtain
\begin{align*}
  &\nabla-\frac{i\eta}{2}\bm{M}=\frac{1}{2}[g^{-1},\bm{\xi}]+L\bm{j}(0),\\
  &\bm{j}(0)=\frac{1}{2}[\nabla\delta(\bm{x})+\delta(\bm{x})\nabla]-\frac{i\eta}{2}\bm{M}\delta(\bm{x}).
\end{align*}
Replace the specific term in \eqref{eqn:app_prob_curr} with the above expression,
\begin{equation}
  \bm{J}'_{\bm{k}}(\eta)=\ointctrclockwise_\mathcal{C}\frac{dz}{2\pi i}\left(\frac{\partial}{\partial t}Tr[\bm{\xi}g]+\frac{1}{2}\frac{\partial}{\partial z}Tr[g\{\bm{\xi},\mathcal{H}\}]
  +LTr[[g,j(0)]g\dot{\mathcal{H}}g]\right).
\end{equation}

The first term turns out to be zero after we take the average of it over time. The second term is zero due to the periodicity of the path. The last term is the only one that contributes. We write it in a more explicit form
\begin{equation}
\begin{split}
  \bm{J}'_{\bm{k}}(\eta)=&L\ointctrclockwise_\mathcal{C}\frac{dz}{2\pi i}\iiint_{-L/2}^{L/2}d\bm{x}d\bm{x}'d\bm{x}''
  [g(\bm{x}',\bm{x})\bm{j}(0)g(\bm{x},\bm{x}'')\dot{\mathcal{H}}(\bm{x}'')g(\bm{x}'',\bm{x}')\\
  &-g(\bm{x}',\bm{x})\bm{j}(0)
  g(\bm{x},\bm{x}'')g(\bm{x}'',\bm{x}')
  \dot{\mathcal{H}}(\bm{x}')]. \label{jgreen}
\end{split}
\end{equation}

The main analogy with Ref.~\cite{NiuThouless-1984} is that the single particle Hamiltonian $h=-\frac{1}{2}(d/dx)^2+U(x,\tau)$  and the Hermitian operator $\mathcal{H}=L_H-i\eta L_A=-\nabla^2+i\eta\bm{M}\cdot\nabla+U'$  are similar. They both possess a kinetic term and a potential term.  In each case, in the deterministic limit, the potential term dominates, consequently the Green functions $g(x,x')$ decays exponentially if $x-x'$ deviates from the peaks.

As  the potential is periodic, the eigenfucntions of  of $\mathcal{H}$  are   Bloch waves, which are superpositions of the Wanner functions, which are localized. The eigenfunctions can be written as
\begin{equation}
 \begin{split}
  \psi_{\bm{r}_0,\bm{k}} =&\sum_{m,n \in\mathds{Z} } e^{i(k_x mL_x+k_y nL_y) }\Gamma_x(x-x_0-mL_x) \Gamma_y(y-y_0-nL_y),
 \end{split}
 \end{equation}
where $\bm{r}_0=(x_0,y_0)$, where $x_0 \in[0,L_x)$ and $y_0 \in[0,L_y)$,  can be regarded  as the band index,  $m$ and $n$ are integers, $\Gamma_x(x-x_0)$ is a localized function peaked at  $x_0$, $\Gamma_y(y-y_0)$ is a localized function peaked at  $y_0$.

Consequently, the Green functions can be calculated as follows
\begin{equation}
  \begin{split}
    g_\alpha(\bm{x},\bm{x'})=&\langle \bm{x}|\frac{1}{z-\mathcal{H}}|\bm{x'}\rangle\\
    =&\iint d^2\bm{r}\langle \bm{x}|\frac{1}{z-\mathcal{H}}
    |\psi_{\bm{r},\bm{\alpha}}\rangle
    \langle\psi_{\bm{r},\bm{\alpha}}
    |\bm{x}'\rangle.
  \end{split}
\end{equation}

In the   deterministic limit or low-temperature limit,  $\mathcal{H} \rightarrow U'$, therefore     \begin{equation}
  \begin{split}
    g_\alpha(\bm{x},\bm{x'})
   \rightarrow &\iint d^2\bm{r}\frac{\psi_{\bm{r},\bm{\alpha}}(\bm{x})
    \psi^*_{\bm{r},\bm{\alpha}}(\bm{x}')}
    {z-U'(\bm{r})}\\
    =&\sum_{m,n,m',n'}e^{ik_x (m-m')L_x+ik_y (n-n')L_y}\iint d^2\bm{r}
    \frac{F(x,x',y,y')}
    {z-U'(\bm{r})}, \end{split}
\end{equation}
   where $F(x,x',y,y')\equiv
    \Gamma_x(x-x_0-mL_x)\Gamma_y(y-y_0-nL_y)
    \Gamma_x(x'-x_0-m'L_x)\Gamma_y(y'-y_0-n'L_y)
  \approx \Gamma_x(x-x'-(m-m')L_x)\Gamma_y(y-y'-(n-n')L_y) $,
  therefore
  \begin{equation}
  \begin{split}
    g_\alpha(\bm{x},\bm{x'})
   \rightarrow
     &\frac{1}{z-U(\bm{x}')}\sum_{m,n,m',n'}e^{ik_x (m-m')L_x+ik_y (n-n')L_y}\Gamma_x(x-x'-(m-m')L_x)
     \Gamma_y(y-y'-(n-n')L_y)
    \\
    =&\frac{\mathcal{N}}{z-U(\bm{x}')}
    \sum_{s,t}e^{ik_x sL_x+ik_y tL_y}
    \Gamma_x(x-x'-sL_x)\Gamma_y(y-y'-tL_y),
    \label{exponential}
  \end{split}
\end{equation}
where $\mathcal{N}$ is the number of different values of $(m,n)$.

It is clear that in the deterministic or low-temperature limit,  the Green functions peak at points with   $|x-x'|=sL_x$ and $|y-y'|=tL_y$, and decay rapidly away from the   the peaks.

On the other hand,   $\bm{j}(0)$ in
\eqref{jgreen} contains Dirac delta functions centred at $\bm{x}=0$, therefore the integrand is considerable only when $\bm{x}=\bm{x}'=\bm{x}''=0$, i.e. $s=t=0$. In this case, in the Green Functions as given in \eqref{exponential},  the $\mathbf{k}$-dependent terms, only appearing as the exponents tend to  vanish, consequently  the Green functions and thus   $\bm{J}'_{\bm{k}}(\eta)$ are  insensitive to $\bm{k}$ in the deterministic or low-temperature limit.

Finally we consider  the analytical continuation of $\eta$ to $i$~\cite{Risken-1585}.  Then $\mathcal{H}= \tilde{\mathcal{O}}$  and $\bm{J}'_{\bm{k}}(i)=\bm{J}_{\bm{k}}$. Since $\bm{J}'_{\bm{k}}(\eta)$ is also insensitive to $\bm{k}$, so is $\bm{J}'_{\bm{k}}(i)$. Hence  we can arrive at the conclusion that $\bm{J}_{\bm{k}}$ is insensitive to $\bm{k}$.

On the other hand, our simulation results confirm topological quantization, hence indirectly confirm the the insensitivity of $\bm{J}_{\bm{k}}$ to $\bm{k}$, consistent with the   validity of the analytic continuation.

\section{Simplification of $\bm{J}_{\bm{k}}^h$}\label{app:simp_Jkh}

After the introduction of the Bloch periodic function, $\bm{J}_{\bm{k}}^h$ can be transformed
\begin{align*}
  \bm{J}_{\bm{k}}^h
  =&\sum_{n\neq0}\frac{\langle e^{i\bm{k}\cdot\bm{r}}v_{0\bm{k}}|
  \left[\nabla+\frac{1}{2}\rho_0^{-1}
  (\hat{\bm{e}}_x\partial_yA-\hat{\bm{e}}_y
  \partial_xA)\right]|e^{i\bm{k}\cdot\bm{r}}w_{n\bm{k}}\rangle
  \langle e^{i\bm{k}\cdot\bm{r}}v_{n\bm{k}}|e^{i\bm{k}\cdot\bm{r}}\dot{w}_{0\bm{k}}\rangle}{E_{0\bm{k}}-E_{n\bm{k}}}\\
  =&\sum_{n\neq0}\frac{\langle v_{0\bm{k}}|
  \left[\nabla+i\bm{k}+\frac{1}{2}
  \rho_0^{-1}(\hat{\bm{e}}_x\partial_yA-
  \hat{\bm{e}}_y\partial_xA)\right]|w_{n\bm{k}}\rangle
  \langle v_{n\bm{k}}|\dot{w}_{0\bm{k}}\rangle}{E_{0\bm{k}}-E_{n\bm{k}}}.
\end{align*}
In the Hilbert space of $w_{n\bm{k}}$, the transformed Fokker-Planck operator must be transformed to
\begin{equation}
  \tilde{\mathcal{O}}'=-(\nabla+i\bm{k})^2-\rho_0^{-1}(\hat{\bm{e}}_x\partial_yA-\hat{\bm{e}}_y\partial_xA)\cdot(\nabla+i\bm{k})+U,
\end{equation}
in order that $\tilde{\mathcal{O}}'w_{n\bm{k}}=E_{n\bm{k}}w_{n\bm{k}}$. Then calculate the derivative of $\tilde{\mathcal{O}}'$ versus $\bm{k}$
\begin{align*}
  &\frac{\partial\tilde{\mathcal{O}}'}{\partial\bm{k}}=-2i(\nabla+i\bm{k})-i\rho_0^{-1}(\hat{\bm{e}}_x\partial_yA-\hat{\bm{e}}_y\partial_xA)\\
  \Rightarrow&\nabla+i\bm{k}+\frac{1}{2}\rho_0^{-1}(\hat{\bm{e}}_x\partial_yA-\hat{\bm{e}}_y\partial_xA)=\frac{i}{2}\frac{\partial\tilde{\mathcal{O}}'}{\partial\bm{k}}.
\end{align*}
Consequently, $\bm{J}_{\bm{k}}^h$ can be simplified,
\begin{equation}
  \bm{J}_{\bm{k}}^h=\frac{i}{2}\sum_{n\neq0}\frac{\langle v_{0\bm{k}}
  |\frac{\partial\tilde{\mathcal{O}}'}{\partial\bm{k}}|w_{n\bm{k}}\rangle}{E_{0\bm{k}}-E_{n\bm{k}}}
  \langle v_{n\bm{k}}|\dot{w}_{0\bm{k}}\rangle.\label{eqn:probability_current_half0}
\end{equation}
After that, calculate the derivative of $\tilde{\mathcal{O}}'w_{n\bm{k}}=E_{n\bm{k}}w_{n\bm{k}}$ versus $\bm{k}$,
\begin{equation*}
  \frac{\partial\tilde{\mathcal{O}}'}{\partial\bm{k}}w_{n\bm{k}}
  +\tilde{\mathcal{O}}'\frac{\partial w_{n\bm{k}}}{\partial\bm{k}}
  =\frac{\partial E_{n\bm{k}}}{\partial\bm{k}}w_{n\bm{k}}
  +E_{n\bm{k}}\frac{\partial w_{n\bm{k}}}{\partial\bm{k}}.
\end{equation*}
Then take the inner product of $v_{0\bm{k}}$ and the above equation,
\begin{equation*}
  \left\langle v_{0\bm{k}}\left|\frac{\partial\tilde{\mathcal{O}}'}{\partial\bm{k}}\right|w_{n\bm{k}}\right\rangle
  +\left\langle\tilde{\mathcal{O}}'^\dag v_{0\bm{k}}\left|\frac{\partial w_{n\bm{k}}}{\partial\bm{k}}\right.\right\rangle
  =\frac{\partial E_{n\bm{k}}}{\partial\bm{k}}\langle v_{0\bm{k}}|w_{n\bm{k}}\rangle
  +E_{n\bm{k}}\left\langle v_{0\bm{k}}\left|\frac{\partial w_{n\bm{k}}}{\partial\bm{k}}\right.\right\rangle.
\end{equation*}

It is straightforward to obtain $\tilde{\mathcal{O}}'^\dag v_{0\bm{k}}=E_{0\bm{k}}^*v_{0\bm{k}}$ and $\langle v_{0\bm{k}}|w_{n\bm{k}}\rangle=0$ for $n\neq0$, which lead to
\begin{equation*}
  \frac{\langle v_{0\bm{k}}|\frac{\partial{\tilde{\mathcal{O}}'}}{\partial\bm{k}}|w_{n\bm{k}}\rangle}{E_{0\bm{k}}-E_{n\bm{k}}}
  =-\left\langle v_{0\bm{k}}\left|\frac{\partial w_{n\bm{k}}}{\partial\bm{k}}\right.\right\rangle=\left\langle\left.\frac{\partial v_{0\bm{k}}}{\partial\bm{k}}\right|w_{n\bm{k}}\right\rangle.
\end{equation*}
By substituting the above equation into \eqref{eqn:probability_current_half0}, one obtains
\begin{equation}
  \bm{J}_{\bm{k}}^h=\frac{i}{2}\sum_{n\neq0}\left\langle\left.\frac{\partial v_{0\bm{k}}}{\partial\bm{k}}\right|w_{n\bm{k}}\right\rangle\langle v_{n\bm{k}}|\dot{w}_{0\bm{k}}\rangle.
\end{equation}
One has a completeness  identity   $\mathbf{1}=
\sum_n|\psi_n\rangle\langle\phi_n|=
\sum_n|w_n\rangle\langle v_n|$ , where the index $\bm{k}$ is omitted for simplicity. Its validity can be justified by the calculation of the matrix elements,
\begin{align*}
  \langle\psi_i|\left[\sum_n|\psi_n\rangle\langle\phi_n|\right]|\psi_j\rangle&=\sum_nT_{in}\delta_{nj}=T_{ij}=\langle\psi_i|\psi_j\rangle,\\
  \langle\phi_i|\left[\sum_n|\psi_n\rangle\langle\phi_n|\right]|\psi_j\rangle&=\sum_n\delta_{in}\delta_{nj}=\delta_{ij}=\langle\phi_i|\psi_j\rangle.
\end{align*}
Therefore, half of the probability current can be simplified further
\begin{equation}
  \begin{split}
    \bm{J}_{\bm{k}}^h=&\frac{i}{2}\langle\partial_{\bm{k}}v_{0\bm{k}}|\partial_t w_{0\bm{k}}\rangle-\frac{i}{2}\langle\partial_{\bm{k}}v_{0\bm{k}}|w_{0\bm{k}}\rangle\langle v_{0\bm{k}}|\partial_t w_{0\bm{k}}\rangle.
  \end{split}
\end{equation}

\end{document}